\newcommand{\figurespath}{.}
\RequirePackage{fix-cm}
\documentclass[twocolumn,epjc3]{svjour3} % EPJC style 
\smartqed  % flush right qed marks, e.g. at end of proof % EPJC style

\RequirePackage{graphicx}

\usepackage{subfigure}
\usepackage{mathrsfs}
\usepackage[T1]{fontenc} % if needed
\usepackage{graphicx}
\usepackage{subfigure}
\usepackage{amsmath}

\usepackage{lineno}

\usepackage{xspace}
\usepackage{placeins}
\usepackage{afterpage}

\usepackage{verbatim}
\usepackage{cite}
\usepackage{url}
\usepackage[hyperindex,breaklinks]{hyperref} 
\usepackage{amssymb}

\usepackage{atlasphysics} % Contains useful shortcuts. Uncomment to use
\usepackage{rotating}
\usepackage{preprintcover}  % load package
\PreprintCoverPaperTitle{Measurement of the top-quark mass in the fully hadronic decay channel from ATLAS data at $\sqrt{s}=7\tev$}  % paper title
\PreprintIdNumber{CERN-PH-EP-2014-208}  % CERN preprint number
\PreprintCoverAbstract{
The mass of the top quark is measured in a data set corresponding to  4.6\,\ifb\ of proton--proton collisions with centre-of-mass energy $\sqrt{s}=7$~Te\kern -0.1em V 
collected by the ATLAS detector at the LHC. 
Events consistent with hadronic decays of top--antitop quark pairs with at least six jets in the final state are selected.
The substantial background from multijet 
production is modelled with data-driven methods that
utilise the number of identified $b$-quark jets and the transverse momentum of the sixth leading jet,
which have minimal correlation.
The top-quark mass is obtained from template fits to the ratio of three-jet to dijet mass. 
The three-jet mass is calculated from the three jets produced in a top-quark decay.
Using these three jets the dijet mass is obtained from the two jets produced in the $W$ boson decay. 
The top-quark mass obtained from this fit is thus less sensitive to the uncertainty in the energy measurement of the jets. 
A binned likelihood fit yields a top-quark mass of
$m_{t} = 175.1 \pm 1.4\stat \pm 1.2\syst$~Ge\kern -0.1em V.
}  % paper abstract
\PreprintJournalName{Eur. Phys. J. C}  % journal name

\journalname{Eur. Phys. J. C} % EPJC style

\begin{document} % EPJC style
%\begin{linenumbers} % EPJC style

\title{Measurement of the top-quark mass in the fully hadronic decay channel from ATLAS data at $\sqrt{s}=7\tev$}

\titlerunning{Measurement of the top-quark mass in the fully hadronic decay channel at $\sqrt{s}=7\tev$}

\author{The ATLAS Collaboration} % \thanksref{e1,addr1}}

\institute{CERN, 1211 Geneva 23, Switzerland \label{addr1}} % EPJC style

\date{Received: date / Accepted: date} % EPJC style

\maketitle % EPJC style

\begin{abstract} % EPJC style
The mass of the top quark is measured in a data set corresponding to  4.6\,\ifb\ of proton--proton collisions with centre-of-mass energy $\sqrt{s}=7$~Te\kern -0.1em V 
collected by the ATLAS detector at the LHC.
Events consistent with hadronic decays of top--antitop quark pairs with at least six jets in the final state are selected.
The substantial background from multijet 
production is modelled with data-driven methods that
utilise the number of identified $b$-quark jets and the transverse momentum of the sixth leading jet,
which have minimal correlation.
The top-quark mass is obtained from template fits to the ratio of three-jet to dijet mass. 
The three-jet mass is calculated from the three jets produced in a top-quark decay.
Using these three jets the dijet mass is obtained from the two jets produced in the $W$ boson decay.
The top-quark mass obtained from this fit is thus less sensitive to the uncertainty in the energy measurement of the jets. 
A binned likelihood fit yields a top-quark mass of
\begin{displaymath} 
    m_{t} = 175.1 \pm 1.4\stat \pm 1.2\syst\gev.
\end{displaymath}
\end{abstract} % EPJC style
\keywords{ATLAS\and LHC\and proton proton collisions\and top quark\and top-quark mass\and fully hadronic}
%%%%%%%%%%%%%%%%%%%%%%%%%%%%%%%%%%%%%%%%%%%%%%%%%%%%%%%%%%%%%

\section{Introduction}
\label{sec-Introduction}
The top quark is the heaviest known fundamental particle and is unique in many
respects. In the Standard Model, its large mass derives 
from a Yukawa coupling to the Higgs 
boson~\cite{bib-HiggsDiscoveryPaper-ATLAS,bib-HiggsDiscoveryPaper-CMS} 
close to unity. Thus it plays a critical role in the quantum corrections to
the electroweak Higgs potential and possible vacuum instability at high
energies (see Ref.~\cite{Sher:1988mj} for a review). 
Because of its large mass, the top quark has a lifetime shorter than the typical time 
scale of hadronisation of coloured quarks to hadrons. Hence, 
the properties of the top quark can be investigated unaffected from non-perturbative 
effects occuring in hadronic bound states. However, the hadronisation of the quarks
and gluons constituting the jets from the decay products of the top quark introduces
an unavoidable sensitivity of the measured top-quark mass on non-perturbative effects.
The top-quark mass $m_{t}$, is also an essential 
parameter in high-precision fits to electroweak observables~\cite{Baak:2012kk}.

The top-quark mass can be determined from decay channels involving hadronic and leptonic decays of
the intermediate $W$ boson.
For the recent world-average top-quark mass value\cite{ATLAS:2014wva}, 
the highest precision~\cite{%
Abazov:2014dpa,
Abazov:2012rp,
Aaltonen:2012va,
Aaltonen:2011dr,
Aaltonen:2011em,
Chatrchyan:2013xza,
Chatrchyan:2013boa,
Chatrchyan:2012ea,
Chatrchyan:2012cz,
ATLAS:2012aj
} 
comes from measurements using the lepton plus jets final state in the decay of top--antitop pairs ($\ttbar$).
This channel has a substantial branching fraction and allows
a relatively unambiguous assignment of jets to partons from the $\ttbar$ decay.
Such events are selected using the lepton and neutrino from the decay of a $W$ boson 
from one member of the top--antitop pair.

Events in which the top--antitop quark pair decays into a fully hadronic final state constitute
both the largest branching fraction and a complementary final state for the determination 
of the top-quark mass. The fully hadronic decay mode has been used in Refs.~\cite{Chatrchyan:2013xza,Aaltonen:2011em} 
to measure the top-quark mass from $\ttbar$ pairs. This decay mode is used in this analysis to measure 
the top-quark mass from $\ttbar$ pairs produced in proton--proton collisions 
provided by the LHC, 
and observed by the ATLAS detector.
The major background to this final state, with orders of magnitude larger cross section, is multijet production 
from proton--proton collisions other than $\ttbar$ pairs. Particular experimental attention is required to precisely
estimate and control this large background. This analysis employs 
a data-driven method to form a multijet background prediction. Selected data events are divided into 
several disjoint regions using two uncorrelated observables, such that $\ttbar$ 
events accumulate only in one of these regions. The background is derived from the 
other regions, determining both the shape and normalisation of 
the background distribution in the signal region.

As the top-quark mass is calculated from the measured energy and momentum of 
re\-con\-struct\-ed jets, an accurate understanding of energy and momentum measurements 
is essential. 
The dependence of the measured top-quark mass on the jet energy measurement uncertainty
is reduced by exploiting the fact that two of the three jets originate from the $W$ boson produced 
in the top-quark decay and that the $W$-boson mass is known very precisely.
The analysis presented in this paper uses the observable $R_{3/2}= m_{jjj}/m_{jj}$ to achieve a cancellation of systematic effects common 
to the masses of the reconstructed top quark ($m_{jjj}$) and associated $W$ boson ($m_{jj}$).

\section{The ATLAS detector}
\label{sec-ATLDet}
The ATLAS detector~\cite{Aad:2008zzm} at the LHC covers nearly the entire solid angle 
around the collision point. The inner detector (ID), which is located closest 
to the interaction point, provides charged-particle tracking in the range of
$|\eta| < 2.5$ where $\eta$ is the pseudorapidity.\footnote{%
ATLAS uses a right-handed coordinate system with its origin at the nominal interaction 
point (IP) in the centre of the detector and the $z$-axis along the beam pipe. The 
$x$-axis points from the IP to the centre of the LHC ring, and the $y$ axis points 
upward. Cylindrical coordinates $(r,\phi)$ are used in the transverse plane, $\phi$ 
being the azimuthal angle around the beam pipe. The pseudorapidity is defined in 
terms of the polar angle $\theta$ as $\eta=-\ln\tan(\theta/2)$. The transverse momentum $\pT$ lies in the $x$--$y$ plane.}
The ID comprises a high\--gra\-nu\-larity silicon pixel detector, a silicon microstrip tracker 
and a transition radiation tracker, and is surrounded by a thin superconducting 
solenoid providing a magnetic field of $2$~T. The electromagnetic and hadronic calorimeters 
are located outside the solenoid and cover the pseudorapidity range $|\eta| < 4.9$. 
Within the region $|\eta|<3.2$, electromagnetic calorimetry 
is provided by barrel and endcap lead/liquid-argon (LAr) sampling calorimeters. 
Ha\-dro\-nic energy measurements are provided by a \allowbreak steel/scin\-tillator tile calorimeter 
in the central region and copper/LAr ca\-lo\-ri\-meters in the endcaps. The forward 
regions are instrumented with copper/LAr and tungsten/LAr calorimeters, %\linebreak
optimised for electromagnetic and ha\-dronic energy measurements, respectively. 
The ca\-lo\-ri\-meter system is surrounded by a muon spectrometer, comprising separate 
trigger and high\--pre\-cision tracking chambers. They measure the deflection of muons 
in a magnetic field with a field integral up to $8$~Tm, generated by one barrel and two 
endcap superconducting air-core toroids.

A three-level trigger system is used. The first-level trigger is implemented in hardware 
and uses a subset of detector information to reduce the event rate to a design value of 
at most $75$~kHz. This is followed by two software-based trigger levels, which together 
reduce the event rate to a few hundred Hz.

The energy scale and resolution of the electromagnetic and hadronic calorimeter systems\cite{Aad:2011he}
as well as the performance of the tracking detector for tagging jets from bottom quarks through 
the displaced decay vertices of $b$-flavoured hadrons\cite{ATLAS-CONF-2011-102,ATLAS-CONF-2012-040,ATLAS-CONF-2012-097}
are of major importance for the precision of this measurement. 
Jet energies measured by the electromagnetic and hadronic calorimeters are adjusted using correction factors, 
obtained from an in situ calibration~\cite{Aad:2011he}, which depend on pseudorapidity ($\eta$) and transverse momentum ($\pT$).

\section{Data, simulation, event selection and reconstruction}
\label{sec-DatSimSel}
\subsection{Data and simulation}
\label{sec-DatSim}
This measurement uses data recorded by the ATLAS detector during 2011 from $7\tev$ proton--proton 
collisions corresponding to an integrated luminosity of $4.6\,\ifb$~\cite{Aad:2013ucp}. 
Events were generated using Monte Carlo (MC) programs in order to
investigate systematic uncertainties, to correct for systematic effects, and 
to generate template distributions used for fitting 
the top-quark mass. 
A fast simulation of the ATLAS detector response,
which is based on full simulation of the tracking detectors and on 
parameterisations for the electromagnetic and hadronic calorimeter showers\cite{ATLAS:2010bfa}, was 
applied to the generated events. For systematic studies a smaller sample of events 
was processed by a full {\sc Geant4}\cite{Agostinelli:2002hh} simulation of the ATLAS 
detector~\cite{Aad:2010ah}. The agreement between parameterised and full simulation was 
verified in detail, as described in Ref.~\cite{ATLAS:2010bfa}. The remaining differences are small and accounted for by a systematic 
uncertainty.  All simulated events were subject to the same selection
criteria and reconstructed using the same algorithms applied to data. 
To generate $\ttbar$ events, the  MC program 
{\sc Powheg-box}\cite{Alioli:2010xd,Frixione:2007nw} 
was employed, which incorporates a theoretical calculation in next-to-leading-order (NLO) accuracy in 
the strong coupling $\alpha_S$, with NLO parton distribution functions (PDFs) 
CT10\cite{Lai:2010vv}. The generated partons are showered
and hadronised by {\sc Pythia}\cite{Pythia}. Adjustable parameters
of {\sc Pythia} are fixed to the values obtained in the {\sc Perugia 2011C} (P2011C)
tune\cite{Skands:2010ak}. Signal events were generated assuming seven different
top-quark mass values from $165.0$ to $180.0\gev$ in steps of $2.5\gev$,
with the largest sample at $172.5\gev$. 
In addition to the hard collisions leading to the $\ttbar$ 
signal, soft scattering processes between the remnants of the protons can take
place. Such processes underlying the signal events are also modelled by 
{\sc Pythia} using the tuned parameters from {\sc Perugia 2011C}. 
Multiple soft proton--proton collisions can take place between different protons in the 
same bunch crossing (in-time pile-up) or arise from collisions in preceding or 
subsequent bunch crossings (out-of-time pile-up) due to the time sensitivity of the detector 
being longer than the time between bunch crossings. 
Such multiple inelastic interactions were also generated by {\sc Pythia}, and are reweighted 
in the simulation to match the distribution of
the number of interactions per bunch crossing measured in the data. %when overlaid to the
This number of interactions ranges
from 3 to 17, with an average of $8.7$.

For studies of systematic uncertainties 
an additional, large sample of signal events was generated at $172.5\gev$, using 
{\sc Powheg-box} and {\sc Pythia} with the {\sc Perugia 2012} tune.

\subsection{Event selection}
\label{sec-Sel}

\begin{table}

\centerline{%
\begin{tabular}{ @{}r@{\hspace*{0.8ex}}p{73mm}@{} }
  \hline
   $\bullet$ & Jet-based trigger \\
   $\bullet$ & $\geq$ 6 jets with $\pT>30\gev$ and $\abseta<2.5$ \\
   $\bullet$ & $\geq$ 5 jets with $\pT>55\gev$ and $\abseta<2.5$ \\
   $\bullet$ & $\Delta R > 0.6$ between pairs of jets with $\pT>30\gev$ \\
   $\bullet$ & Jet vertex fraction JVF\,$>0.75$\\
   $\bullet$ & Reject events w.\ isolated electrons with $\ET>25\gev$ \\
   $\bullet$ & Reject events w.\ isolated muons with $\pT>20\gev$ \\
   $\bullet$ & Exactly 2 $b$-tagged jets among the four leading jets\\
   $\bullet$ & Missing transverse momentum significance \newline 
               $\met [\mathrm{Ge\kern -0.1em V}]/\sqrt{\HT [\mathrm{Ge\kern -0.1em V}]}<3$ \\
   $\bullet$ & Centrality ${\cal C}>0.6$ \\
  \hline
\end{tabular}}%
\caption{\label{tab:evSel}
Summary of event selection requirements for signal events.%
}
\end{table}

A jet-based trigger is used in which the jets are reconstructed 
in the online trigger 
system~\cite{Aad:2012xs}. This jet reconstruction executes the anti-$k_t$ jet algorithm\cite{Cacciari:2008gp} 
with a radius parameter of $0.4$ using clusters of energy deposition in adjacent calorimeter 
cells (topological clusters)~\cite{Lampl:1099735,ATLAS-CONF-2012-20}. 
At least five jets with a nominal $\pT$ threshold of $30\gev$ are required to trigger and
record an event.

Events are selected according to the requirements listed in Table~\ref{tab:evSel} and 
detailed in the following.
Only events with a well-reconstructed primary vertex formed by at least five tracks
with $\pT>400\mev$ per track
are considered for the analysis, where the primary vertex is the reconstructed
vertex with the highest summed $\pT^2$ of associated tracks.
Similar to the online trigger system, jets are reconstructed offline by
the anti-$k_t$ jet algorithm with a radius parameter of $0.4$ using topological clusters. 
The jet energies are calibrated following Refs.~\cite{Aad:2014bia,ATLAS-CONF-2013-004,ATLAS-CONF-2013-002}.
For the parameterised simulation a dedicated jet energy calibration is used which is
obtained in the same manner as for the full simulation. 
To ensure that events selected by the trigger are on the plateau of the efficiency curve, 
only events which have at least five jets, each with $\pT>55\gev$,
and $\Delta R > 0.6$ \footnote{Distances between 
particles or jets are measured using $\Delta R = \sqrt{(\Delta\phi)^2 + (\Delta\eta)^2}$ 
where $\Delta\phi$ and $\Delta\eta$ are the differences in $\phi$ and $\eta$ between 
the two objects.} between every pair of jets with $\pT>30\gev$ 
are considered. 
The measured trigger efficiency of $90\%$ agrees with the expectation from 
simulation to within 5\%. This remaining difference is considered as a source of 
systematic uncertainty in Sect.~\ref{sec-systematics}.

A signal event is required to have at least six jets. 
Only jets in the central part of the calorimeter 
($\abseta<2.5$) and with $\pT> 30\gev$ are considered for the $\ttbar$ mass analysis,
but for the background determination the sixth leading jet has a looser requirement of $\pT> 25\gev$.
For a jet to be considered, at least 75\% of its summed track $\pT$ must be due to tracks coming from the primary vertex 
(jet vertex fraction JVF\,$>0.75$). 
Jets in an event are rejected if an identified electron 
is closer than $\Delta R=0.2$. 

Events with identified isolated electrons with $\ET>25\gev$ or muons with $\pT>20\gev$ 
are rejected. Details of the lepton identification are given in Refs.~\cite{CERN-PH-EP-2011-117,Aad:2014zya}. 
Events are kept for further analysis when at most two of the four leading 
transverse momentum jets are 
identified as $b$-tagged jets by a neural network trained on decay vertex properties. The neural network provides an 
identification efficiency of 70\% for jets from $b$-quarks, a rejection factor 
of about 130 for jets arising from light partons, and a factor of about 5 for jets arising from 
$c$-quarks~\cite{ATLAS-CONF-2012-043}. 
In the signal region, exactly two of the four leading 
transverse momentum jets are required to be $b$-tagged by the neural network. 
Events with mismeasured jet energies or with potential leptonic decays that include neutrinos are 
removed by requiring a missing transverse momentum significance
$\met [\mathrm{Ge\kern -0.1em V}]/\sqrt{\HT [\mathrm{Ge\kern -0.1em V}]}$ 
of less than 3. Here $\HT$ is the scalar sum of the transverse momenta of all selected jets in the 
event. The $\met$ is obtained as in Ref.~\cite{ATLAS:2012aj} as the magnitude of the negative 
vectorial sum of calorimeter energy deposits projected onto the transverse plane, 
plus the transverse momenta of identified muons measured by the tracking detector and muon spec\-tro\-meter. 
Measured energy deposits in the ca\-lo\-ri\-meters are corrected according to
the identified object (high-$\pT$ jet, photon, electron, muon); otherwise 
energy deposits are calibrated with the local ha\-dron\-ic calibration scheme detailed in 
Ref.~\cite{Barillari:2009zza}.
The contribution from multijet background events is reduced by using the centrality ${\cal C}$
of the signal events, which is different from the value in multijet events due to the large top-quark mass.
Events are required to have ${\cal C}> 0.6$, with
\begin{equation}
   {\cal C} = \frac{\sum_j^{\mathrm{jets}}{E_{{\mathrm{T}},j}}}{\sqrt{\left(\sum_j^{\mathrm{jets}}p_j\right)^2}}\ \ ,
\end{equation}
where $E_{{\mathrm{T}},j}$ is the scalar transverse energy 
and $p_j=(E_j,\vec{p}_j)$ the four-momentum of the $j^{\mathrm{th}}$ selected jet,
and the sum is over all selected jets.

\subsection{Reconstruction}
\label{sec-Reco}
\begin{figure}[t]
\label{fig:logLike}
\centerline{\includegraphics[width=0.49\textwidth]{\figurespath/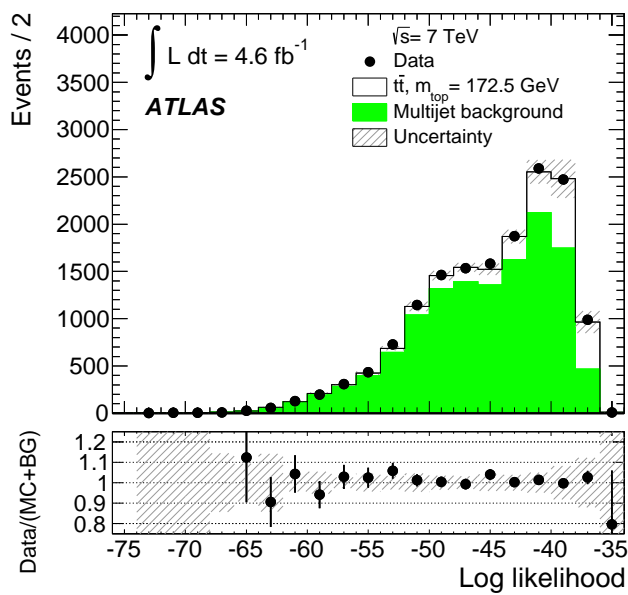}}
\caption{\label{fig:loglike}%
Comparison of the distribution of the unnormalised logarithmic likelihood for the reconstruction of
fully hadronic $\ttbar$ events in the data with 
expectations for a top-quark mass value of $172.5\gev$. The graph in the lower inset shows the ratio
of data to the sum of $\ttbar$ MC signal and the modelled multijet
background (see Sect.~\ref{sec-Background}). The error bars indicate the statistical
uncertainty of the data. The shaded bands show the 
statistical and systematic 
(see Sect.~\ref{sec-systematics}) uncertainty on the 
expected signal and background distributions.
}%
\end{figure}
In each selected event, a fully hadronic $\ttbar$ final state is reconstructed using 
the six or more jets.
In order to achieve this, the jets in data are assigned to the decay partons expected from the
decay of the top quark and the related intermediate $W$ boson, assuming a leading-order decay.
Exploiting the knowledge 
of the precisely known mass of the $W$ boson and the Breit--Wigner lineshapes 
of the top quark and the $W$ boson decay, a kinematic fit~\cite{Erdmann:2013rxa} based on a likelihood 
function similar to the one described in Ref.~\cite{ATLAS:2012aj} assists in establishing the assignment of reconstructed 
jets to partons. The fit is
performed maximising the logarithmic likelihood, defined as the product of Breit-Wigner
distributions for the two top-quark and $W$ boson masses, and MC derived transfer functions
for each of the six jets.
The Breit--Wigner lineshape functions use the world-average values of the 
$W$ boson mass ($80.4\gev$) and decay width ($2.1\gev$) from Ref.~\cite{PDG}.
The masses of the top quark and antiquark are assumed to be equal for the Breit--Wigner 
lineshape and free to float in
the fit. The top decay 
width is kept fixed at $1.3\gev$, corresponding to a top-quark mass of $172.5\gev$.
The energies of the partons are transferred to the measured jet energies by transfer functions 
derived from simulation and parameterised by superpositions of two Gaussian
functions.
It is required, furthermore, that the fit assigns the 
$b$ and $\overline{b}$ quarks from the $\ttbar$ decay to any two of the four leading jets.
Maximising the logarithmic likelihood establishes the best assignment of
reconstructed jets to partons from the $\ttbar$ decay. Figure~\ref{fig:loglike}
shows the distribution of the unnormalised logarithmic likelihood value obtained per
event and compared with the Monte Carlo prediction of the $\ttbar$ signal added
to the modelled multijet background (see Sect.~\ref{sec-Background}). 
The prediction is in good agreement with the shape of the distribution.
Requiring 
the logarithmic likelihood value to be greater than $-45$ 
removes events which yield a low probability under a $\ttbar$ decay hypothesis.
The cut rejects about 47\% of the multijet background events, while 79\% of 
the fully hadronically decaying $\ttbar$ events pass the cut.

After applying the above selection requirements and performing the $\ttbar$ reconstruction
$15\thinspace551$ events remain in the signal region for the measurement of the top-quark mass 
(see Table~\ref{table:signalPres2}).
The expected fraction of $\ttbar$ events in this region
without any restriction on $R_{3/2}$
is about $17\%$, corresponding to a selection efficiency of $\approx$~0.5\%.

\section{Modelling of multijet background}
\label{sec-Background}
The multijet background contribution is large and cannot be removed completely from any 
distribution used to measure the top-quark mass in the fully hadronic final state. 
Currently only leading-order theory 
calculations for final states with up to six parton are available in MC
generator programs.
Therefore, the multijet background is determined from the data. 

For this approach,
selected events are divided into six regions ($A$--$F$) by using 
two observables with minimal correlation: the number of $b$-tagged jets 
and the transverse momentum of the sixth leading jet, $\pt^{\mathrm{6th\ jet}}$. 
The correlation in $\ttbar$ events is estimated in simulation to be $\rho=0.009$.
The six regions, defined by three bins of the number of $b$-tagged jets 
and two ranges in $\pt^{\mathrm{6th\ jet}}$, are detailed in
Table~\ref{table:signalPres2}.
Region $F$, which is the signal region, i.e.\ two $b$-tagged jets with 
$\pt^{\mathrm{6th\ jet}}>30\gev$, contains the largest fraction of
$\ttbar$ events in addition to multijet background events. 
\begin{table*}
\renewcommand{\arraystretch}{1.9}

\centerline{%
\begin{tabular}{|c||c|r|r||c|r|r|}
\cline{2-7}
\multicolumn{1}{l}{~}
         & \multicolumn{3}{|c||}{ $\pT^{\mathrm{6th\ jet}}\leq30\gev$}
         & \multicolumn{3}{|c|}{  $\pT^{\mathrm{6th\ jet}}>   30\gev$}
\\ \cline{2-7}
\multicolumn{1}{l|}{~}
         &  \raisebox{0ex}[5mm][0mm]{~}      & \multicolumn{1}{|p{16mm}|}{Data events  $N_R^{\mathrm{obs}}$}
                  & \multicolumn{1}{|p{18mm}||}{Signal MC events  $N_R^{\mathrm{sig}}$}
         &        & \multicolumn{1}{|p{16mm}|}{Data events  $N_R^{\mathrm{obs}}$}
                  & \multicolumn{1}{|p{18mm}|}{Signal MC events  $N_R^{\mathrm{sig}}$}
\\ \cline{1-1}\cline{3-4}\cline{6-7}
$b$-tagged jets
         &  \raisebox{-0.5ex}[2mm][0mm]{\rotatebox{90}{Region $R$}} 
                  & \multicolumn{2}{|c||}{signal fraction}
         &  \raisebox{-0.5ex}[2mm][0mm]{\rotatebox{90}{Region $R$}} 
                  & \multicolumn{2}{|c|}{signal fraction}
\\ \hline\hline
$0$ & $A$ &   $93,732$ & $ 306\pm 4$%
    & $B$ &  $286,416$ & $2607\pm11$%
\\ \cline{3-4}\cline{6-7}
    &     &  \multicolumn{2}{|c||}{           $0.33\pm0.01\%$}
    &     &  \multicolumn{2}{|c|} {$\phantom{1}0.91\pm0.01\%$}
\\ \hline\hline
$1$ & $C$ &   $23,536$ & $ 678\pm 5$%
    & $D$ &  $ 77,301$ & $5117\pm14$%
\\ \cline{3-4}\cline{6-7}
    &     &  \multicolumn{2}{|c||}{           $2.88\pm0.04\%$}
    &     &  \multicolumn{2}{|c|} {$\phantom{1}6.62\pm0.04\%$}
\\ \hline\hline
$2$ & $E$ &   $ 4,532$ & $ 399\pm 5$%
    & $F$ &  $ 15,551$ & $2582\pm13$%
\\ \cline{3-4}\cline{6-7}
    &     &  \multicolumn{2}{|c||}{$           8.80\pm0.29\%$}
    &     &  \multicolumn{2}{|c|} {$          16.60\pm0.27\%$}
\\ \hline
\end{tabular}
}%
\caption{\label{table:signalPres2}%
         Event yields for the six regions, defined by the number of $b$-tagged jets and the 
         transverse momentum of the sixth leading jet $\pt^{\mathrm{6th\ jet}}$, are listed for data and 
         $\ttbar$ simulation assuming $m_{t}=172.5\gev$ with statistical uncertainty. 
         The $\ttbar$ fractions are 
         derived from the observed numbers of events and their statistical uncertainties.}
\end{table*}
\begin{figure*}%[h]
\hspace*{-5mm}
\subfigure{%
\label{fig:mjjc}
\includegraphics[width=0.34\textwidth]{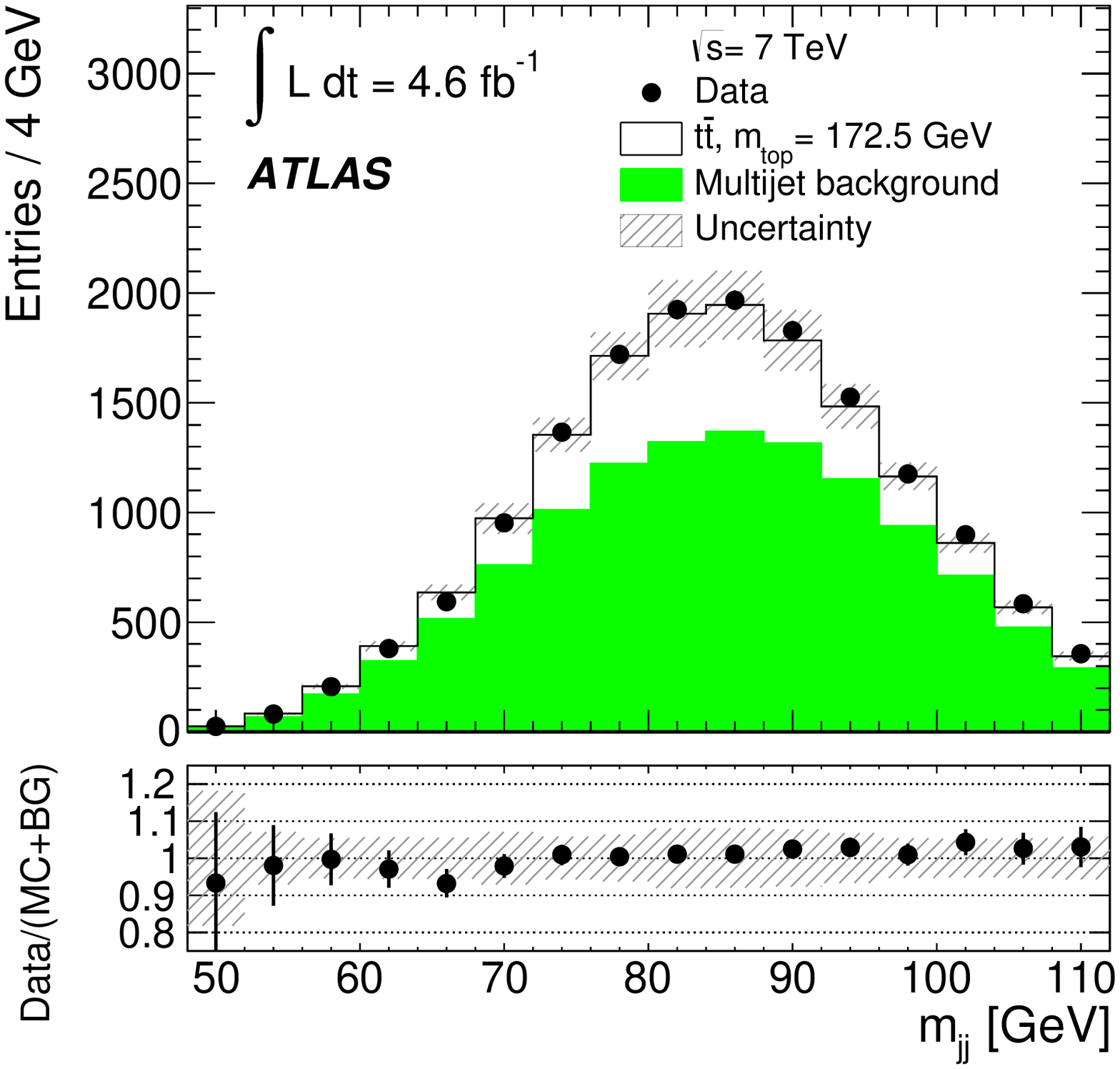}
}%
\subfigure{%
\label{fig:mjjjc}
\includegraphics[width=0.34\textwidth]{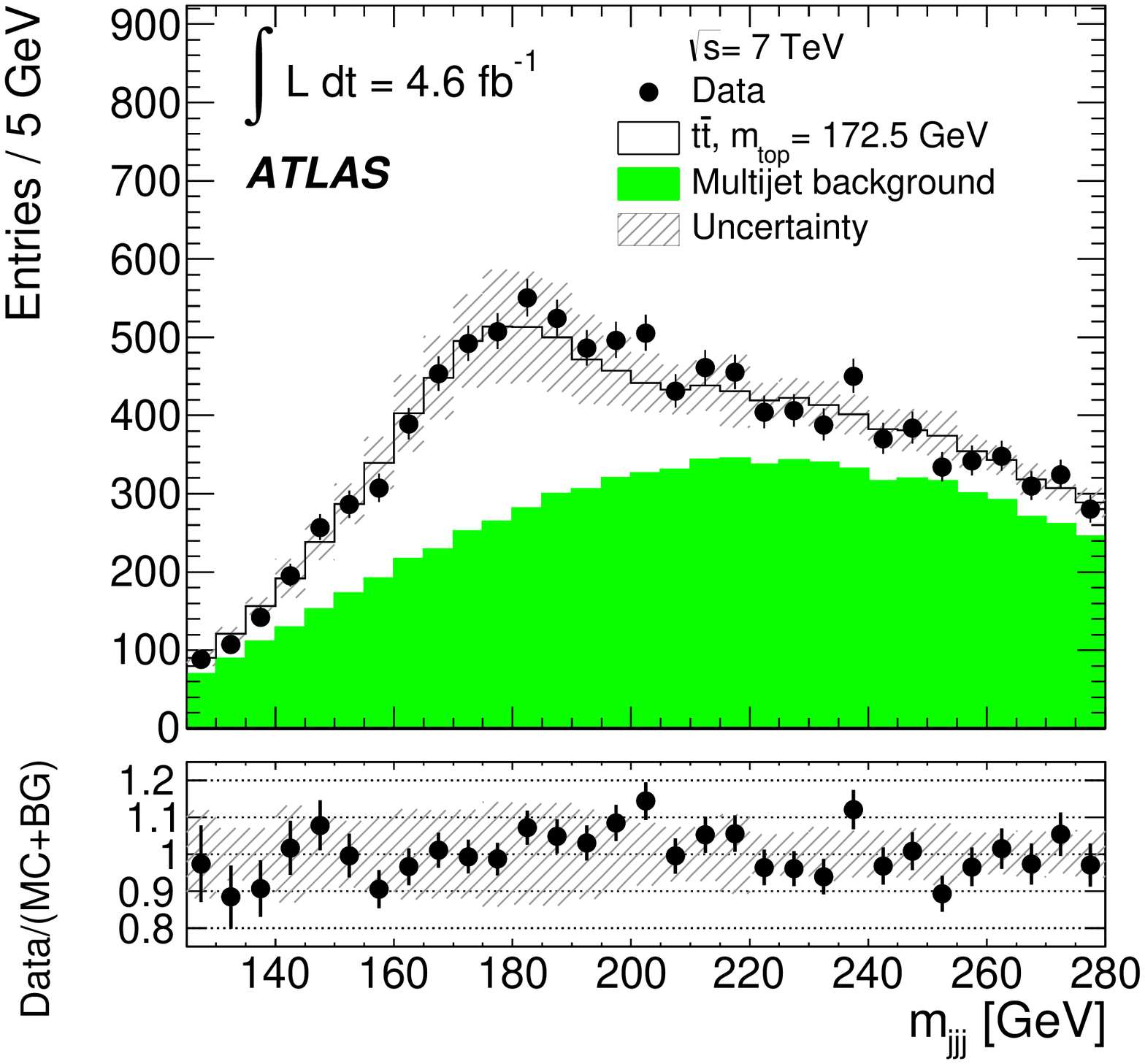}
}%
\subfigure{%
\label{fig:R32c}
\includegraphics[width=0.34\textwidth]{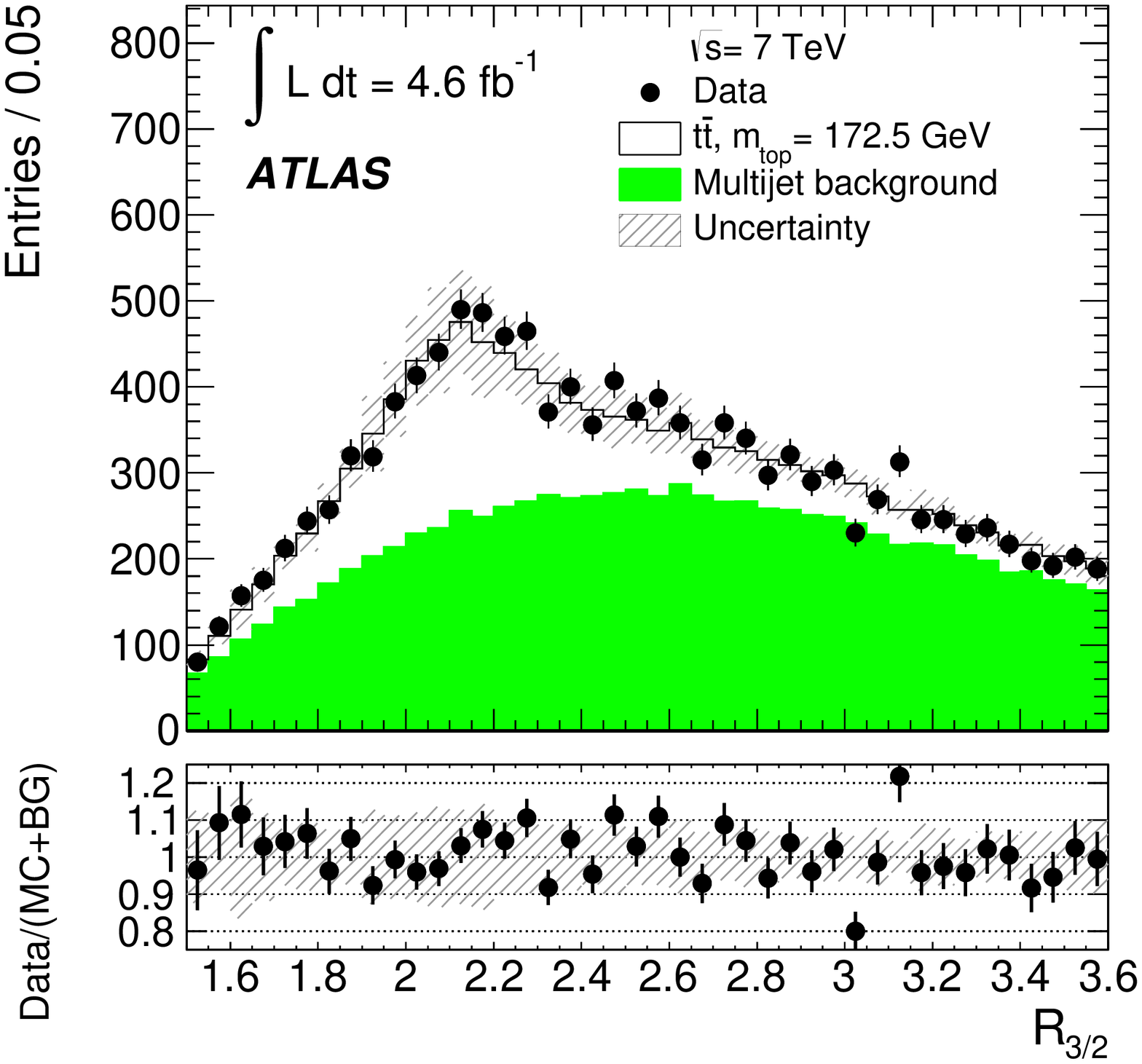}
}%
\caption{\label{fig:recoc}%
Distributions of (left) dijet mass $m_{jj}$, (middle) three-jet mass $m_{jjj}$, 
and (right) ratio of three-jet mass to dijet mass $R_{3/2}$, measured in data and compared to expectations after 
applying all analysis event selection criteria (i.e.\ for region $F$). The shape and normalisation of the
multijet background distributions (green shaded histograms) are calculated using 
Eq.~(\ref{eqn-ABCDEF}).
The distributions for the $\ttbar$ events (white histograms) are taken from 
the MC simulation using a top-quark mass value of $172.5\gev$.
The insets under the distributions show the ratio of data to the summed contributions
of $\ttbar$ MC signal and modelled multijet background (see Sect.~\ref{sec-Background}). The 
error bars represent the statistical uncertainties on the data. 
The shaded bands show the 
statistical and systematic 
(see Sect.~\ref{sec-systematics}) uncertainty on the 
expected signal and background distributions.
}%
\end{figure*}
Regions $A$ through $E$ are depleted in $\ttbar$ events, but enhanced in multijet 
background events. 
The data yields in these regions ($N_R^{\mathrm{obs}}$, $R=A, \ldots, E$) and 
the expected number of $\ttbar$ events from MC simulation, $N_R^{\mathrm{sig}}$, using $m_t=172.5\gev$  
are listed in 
Table~\ref{table:signalPres2}. The table also quotes the derived fraction $N_R^{\mathrm{sig}}/N_R^{\mathrm{obs}}$ 
of $\ttbar$ events in the respective region.
The $\ttbar$ event fraction in each region other 
than $F$ is accounted for by subtracting from data, $N_R^{\mathrm{obs}}$, the 
number of $\ttbar$ events predicted by the MC simulation, $N_R^{\mathrm{sig}}$, for a 
top-quark mass value of $175\gev$:
\begin{equation}
\label{eqn-top_bkg_subtraction}
  N_R^{\mathrm{bkg}} = N_R^{\mathrm{obs}} - N_R^{\mathrm{sig}}
\end{equation}
for region $R=A, \ldots,  E$. Due to the small $\ttbar$ fractions in region $A$ to $E$,  
the top-quark mass value chosen in the simulation used for this subtraction 
procedure marginally affects the value of $m_{t}$ measured in this analysis. Therefore,
the value of $m_{t}$ closest to the measured value (see Sect.~\ref{sec-TopMass}) is 
used in the simulation for subtraction.
The small dependence on the $\ttbar$ MC simulation 
introduced by this subtraction is accounted for by a 
systematic uncertainty (see Sect.~\ref{sec-BackgroundSystematics}).

Given the tiny correlation of $0.9\%$ predicted by MC simulation studies for the two observables 
used to define the regions, the total 
number of multijet background events, $N_F^{\mathrm{bkg}}$, 
in region $F$ can be estimated by cross-multiplication, for example, from the ratio of the number of events in region
$B$ to region $A$ scaled by the number of events in region $E$. 
To obtain the distribution of multijet background events, $N_F^{\mathrm{bkg}}(x)$, for any given observable $x$ (e.g. $R_{3/2}$)
to the distribution in region $F$
either of the following formulae can be used:
\begin{eqnarray}
\label{eqn-ABEF-CDEF}
N_F^{\mathrm{bkg}}(x) = N_E^{\mathrm{bkg}}\cdot\frac{N_B^{\mathrm{bkg}}(x)}{N_A^{\mathrm{bkg}}}\ \ 
\ \ \ \ &\!{\mathrm{or}}\!& \ \ \ \ 
\\ \nonumber
N_F^{\mathrm{bkg}}(x) = N_E^{\mathrm{bkg}}\cdot\frac{N_D^{\mathrm{bkg}}(x)}{N_C^{\mathrm{bkg}}}
,
\ \ \ \ &\!{\phantom{\mathrm{or}}}\!& \ \ \ \ 
\end{eqnarray}
hence
\begin{eqnarray}
\label{eqn-ABCDEF}
N_F^{\mathrm{bkg}}(x)=\frac{N_E^{\mathrm{bkg}}}{2}&\!\!\cdot\!\!&\left(\frac{N_B^{\mathrm{bkg}}(x)}{N_A^{\mathrm{bkg}}}+\frac{N_D^{\mathrm{bkg}}(x)}{N_C^{\mathrm{bkg}}}\right)
.
\end{eqnarray}
Here, 
$N_B^{\mathrm{bkg}}(x)$ and $N_D^{\mathrm{bkg}}(x)$ define 
the shape of the distributions for an observable $x$, while 
the appropriate normalisation is achieved by scaling with
the total number of events ($N_A^{\mathrm{bkg}}$, $N_C^{\mathrm{bkg}}$, $N_E^{\mathrm{bkg}}$) in the 
respective region. 
Equation~(\ref{eqn-ABCDEF}) is
used to determine the multijet background while Eqs.~(\ref{eqn-ABEF-CDEF}) are used to 
estimate the systematic uncertainties on the modelled background (see Sect.~\ref{sec-BackgroundSystematics}).

Figure~\ref{fig:recoc} shows the distributions of the dijet mass,
the three-jet mass, and their ratio, $R_{3/2}=m_{jjj}/m_{jj}$, 
after applying the event selection and jet assignments detailed in Sect.~\ref{sec-DatSimSel}. 
In calculating $R_{3/2}$ values for an event, $m_{jjj}$ of both top-quark candidates
and $m_{jj}$ of the related $W$ boson candidate are considered.
Superimposed in Fig.~\ref{fig:recoc}
is the sum of the distributions for the $\ttbar$ events obtained
from MC simulation using $m_t=172.5\gev$ plus the multijet
background estimated using Eq.~(\ref{eqn-ABCDEF}). The distributions of the ratios of
data to the sum of the signal MC events plus background model seen in 
Fig.~\ref{fig:recoc} show that the data-driven approach yields a 
reliable model of the multijet background.

\section{Top-quark mass measurement}
\label{sec-TopMass}
The top-quark mass is obtained from a binned likelihood fit to the $R_{3/2}$
distribution shown in Fig.~\ref{fig:recoc}. 
As noted above, two values of $R_{3/2}$ are contributed by each event, reconstructed separately from the 
top and antitop-quark candidates.
Because equal masses are assumed 
for the Breit--Wigner lineshapes for 
the top quark and antiquark in the kinematic fit for the jet assignments, the two values are correlated
at the level of approximately 60\% according to MC simulation. 
This is corrected for in the statistical treatment described below.
Templates are created for both the simulated top-quark contribution
to the $R_{3/2}$ distribution and the modelled background distribution. The top-quark contribution
is parameterised by the sum of a Gaussian function and a Landau function which account, respectively, for 
the correctly reconstructed top-quark events and for the combinatorial background due to 
mis-assignment of jets to partons (see Sect.~\ref{sec-DatSimSel}). 
This description involves six parameters. 

\begin{figure*}%[t]
\centerline{%
  \includegraphics[width=0.83\textwidth]{\figurespath/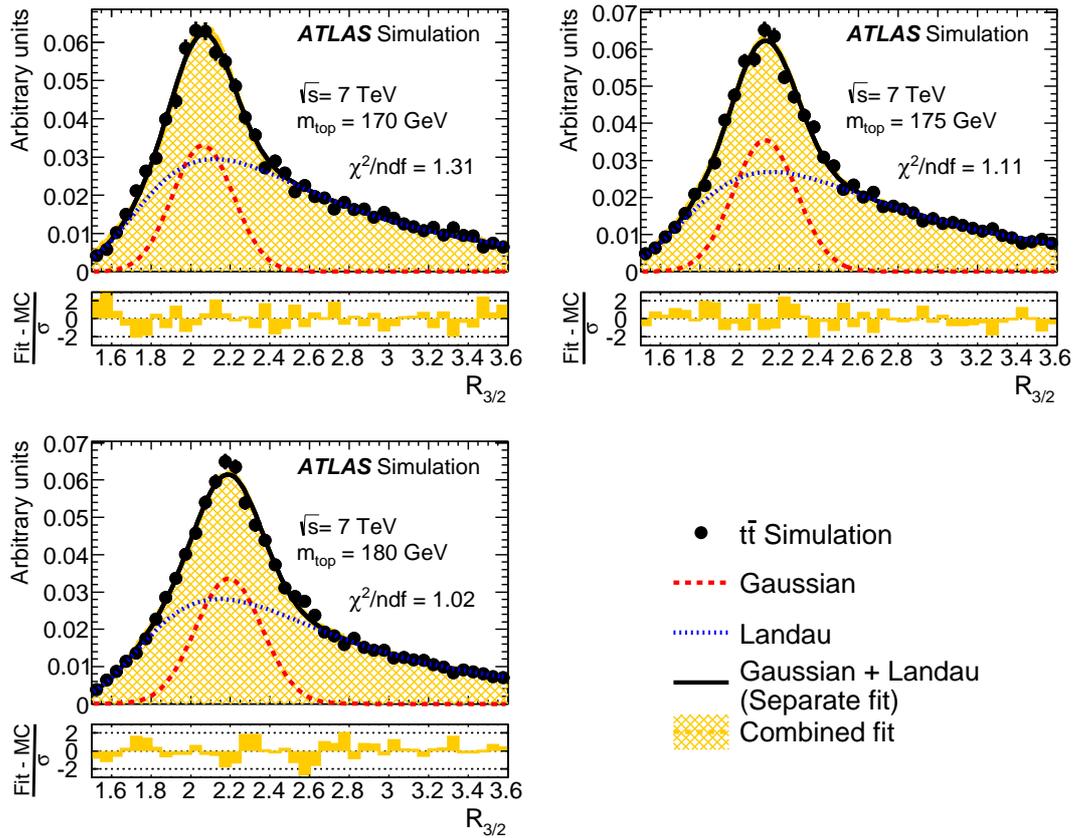}%
}%
   \caption{\label{fig:sigSep}%
            Templates for the $R_{3/2}$ distribution for $\ttbar$ MC 
            simulation using top-quark mass values of 
            $170.0,\ 175.0$ and $180.0\gev$,
            respectively. For each top-quark mass, the $R_{3/2}$ distribution is fitted by the sum (black solid) of a 
            Gaussian (red dashed) and Landau (blue dotted) function. Superimposed (orange cross-hatched) are the 
            templates obtained from
            a combined fit of all $R_{3/2}$ distributions using a linear dependence of parameters of the 
            Gaussian and Landau functions on the top-quark mass value. 
            The insets under the distributions show the difference Fit$-$MC between the combined fit and the simulated $R_{3/2}$
            histogram normalised to the statistical uncertainty $\sigma$ of the corresponding $R_{3/2}$ bin.
           }
\end{figure*}
A two-step approach is used to obtain an 
$m_{t}$-dependent representation of the templates. Firstly,
the $R_{3/2}$ distribution from each of the seven simulation samples of different $m_{t}$ is fitted separately
to determine the six parameters for each template mass. This yields a good description of the $R_{3/2}$
distributions per chosen $m_t$ (see Fig.~\ref{fig:sigSep}). MC simulation has shown that each 
of the six parameters of the Gaussian and Landau functions depend linearly 
on the input top-quark mass. Secondly, from the parameter values obtained by these separate fits,
initial values for offsets and slopes of the linear $m_t$ dependencies are derived 
and then used as inputs to a combined, simultaneous fit to all seven $R_{3/2}$ distributions. 
In total 12 parameters are determined by the combined fit,
which yields a $\chi^2$ per number of degrees of freedom (ndf) of $\chi^2/\mathrm{ndf}=298/282=1.06$. 
Both the individual and the combined fit results are shown for three of the seven $m_t$ values in 
Fig.~\ref{fig:sigSep}.

The modelled multijet background, obtained using Eq.~(\ref{eqn-ABCDEF}), is parameterised by a Gaussian function plus a linear
function, thus involving five parameters. The resulting fit to data is shown in Fig.~\ref{fig:backSep} and yields 
$\chi^2/$ndf$=40/36=1.08$. The shape of 
the fitted parameterisation is assumed to be independent of the top-quark mass while the normalisation 
is obtained from fitting to the data distribution. Any residual dependence of this parameterisation
on the top-quark mass is accounted for by a systematic uncertainty (see Sect.~\ref{sec-systematics}).

The $R_{3/2}$ distribution is fitted for the top-quark mass using the templates for both the top-quark signal 
and the modelled multijet background distribution described above. Defining the likelihood function as a product of Poisson probabilities
\begin{equation}
\label{equ:bLike}
  \mathcal{L}(R_{3/2}|m_{t})= \prod_{j}^{\mathrm{bins}} \left( \frac{\lambda_{j}^{N_{F,j}^{\mathrm{obs}}}}{N_{F,j}^{\mathrm{obs}}!} 
                                              \right)\exp({-\lambda_{j}})
,
\end{equation}
a binned likelihood fit is applied.
For the $R_{3/2, j}$, i.e.\ the $j^{\mathrm{th}}$ bin of the $R_{3/2}$ distribution, 
$N_{F,j}^{\mathrm{obs}}\equiv N_F^{\mathrm{obs}}(R_{3/2, j})$ and $\lambda_{j}$ are the observed and expected number of 
events in that bin. Here, the expected number of events in a bin is given by the sum of $\ttbar$ events
$N_{F,j}^{\mathrm{sig}}(m_{t})$, as derived from the signal templates, and multijet background events 
$N_{F,j}^{\mathrm{bkg}}\equiv N_F^{\mathrm{bkg}}(R_{3/2, j})$,
\begin{equation}
 \lambda_{j} = (1-f_{\mathrm{bkg}}) N_{F,j}^{\mathrm{sig}}(m_{t}) + f_{\mathrm{bkg}} N_{F,j}^{\mathrm{bkg}},
\end{equation}
where $f_{\mathrm{bkg}}$ is the fraction of multijet background events, which is determined by the fit. 

Equation~(\ref{equ:bLike}) is maximised with respect to $m_{t}$ and $f_{\mathrm{bkg}}$ for
$R_{3/2}$ values between $1.5$ and $3.6$, 
taking the normalisation from data, yielding 
\begin{equation}
\label{eqn-fitresult}
     m_{t} = 175.06 \pm 1.35\stat\gev
\end{equation}
for a background fraction of 
 $f_{\mathrm{bkg}}=0.72  \pm 0.01$ 
and 
 $\chi^2/$ndf $=48/39=1.23$. The difference between the fitted background fraction and the value quoted in 
 Sec.~\ref{sec-Reco} is due to the restricted $R_{3/2}$ range used in the fit.
The result of this fit is shown in Fig.~\ref{fig:bestFitData}. 
The $\chi^2/$ndf value is enlarged by the statistical correlation between the two 
$R_{3/2}$ values from each event. Its impact has been incorporated
in the quoted statistical uncertainty\footnote{The uncorrected statistical uncertainty obtained from the fit yields 1.15 GeV.} of Eq.~(\ref{eqn-fitresult}) as follows.
\begin{figure}%[b]
\centerline{%
  \includegraphics[width=0.5\textwidth]{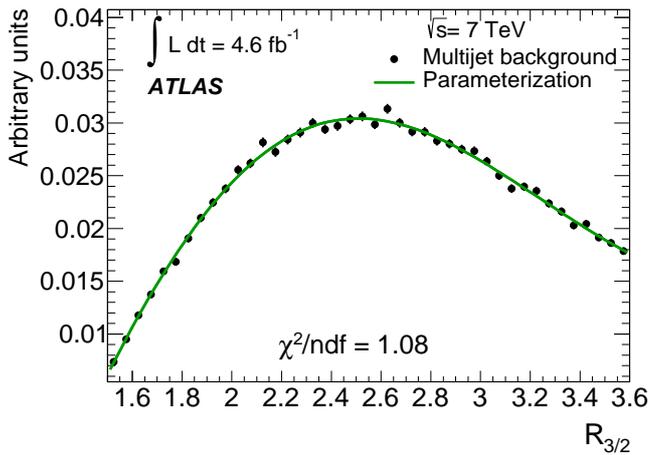}% 
}%
   \caption{\label{fig:backSep}%
           Distribution of $R_{3/2}$ for multijet background events according to 
           the data-driven prescription of Eq.~(\ref{eqn-ABCDEF}), normalised to unit integral. The
           parameterisation of the distribution by the sum of a Gaussian function and 
           a linear function is superimposed.
   }
\end{figure}
\begin{figure}%[h]
\centerline{%
   \includegraphics[width=0.5\textwidth]{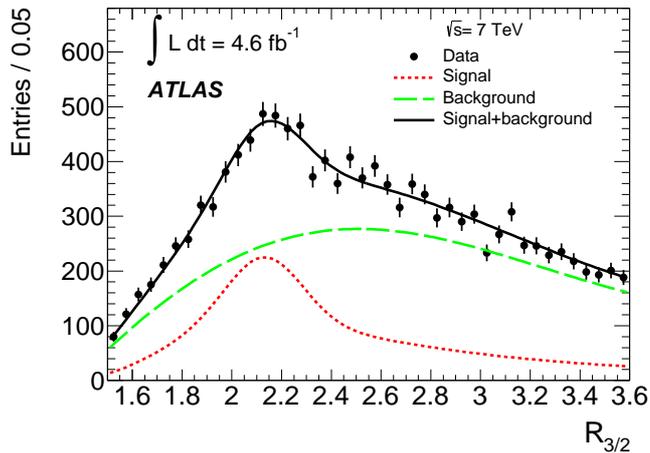}%
}%
    \caption{\label{fig:bestFitData}%
       Result of the fit of Eq.~(\ref{equ:bLike}) (solid black) to the measured $R_{3/2}$ 
       distribution. The red dotted curve shows the contribution from top-quark events and corresponds to the black curve in Fig.~\ref{fig:sigSep};
       the green dashed line is the modelled multijet background.
       }
\end{figure}
\begin{figure}%[h]
\centerline{%
\includegraphics[width=0.5\textwidth]{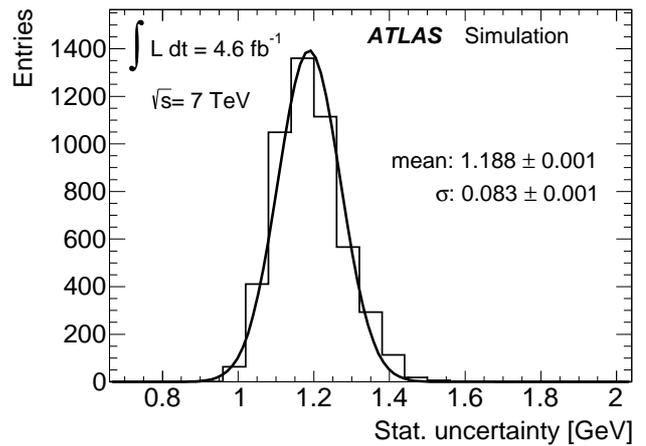}%
}%
\caption{\label{fig:massErrorR32}%
      Expected statistical uncertainty on the top-quark mass obtained from 5000~pseudo-experiments using 
      $\ttbar$ MC simulation events assuming $m_{t}=175\gev$ and neglecting correlations between the
      two $R_{3/2}$ values per event.
 }%
\end{figure}

The statistical uncertainty of the fit is studied by performing
pseudo-experiments, where 5000 pseudo-datasets of $R_{3/2}$ values, each statistically
equivalent to the data, are assembled from values
randomly picked from signal and background histograms\footnote{The signal histograms used to draw pseudodata include $\approx$ 45000 events for the 172.5 GeV mass point sample and 4500-6500 events for the remaining mass points. 
The background histogram is derived using $\approx$ 230000 data events in the control regions defined in Sec.~\ref{sec-Background}.}. They are obtained from $\ttbar$ MC simulation\footnote{%
A single event may be used several times in
different data sets. The correlation introduced by this resampling technique is 
corrected in all distributions and results presented in this paper as described 
in Ref.~\cite{MAN-HEP-99-4}.} generated for $m_{t}=175\gev$, and from the multijet
background estimate, detailed in Sec.~\ref{sec-Background}, respectively.
Pseudo-datasets are created from two-dimensional histograms for the full MC sample of 
$R_{3/2}$ from the top-quark candidate versus $R_{3/2}$ of the
top-antiquark candidate in an event, thereby accounting for the $60\%$ correlation. 
Similarly, one-dimensional histograms are used to produce pseudo-datasets which do not include the correlations. 
The top quark mass and its statistical uncertainty are evaluated for each pseudo-dataset, using
the likelihood fit of Eq.~(\ref{equ:bLike})

The expected statistical uncertainty of the fit when neglecting the correlation 
is shown in Fig.~\ref{fig:massErrorR32}. 
A fit of a Gaussian function to the output of the 5000 pseudo-experiments yields an expected
statistical uncertainty of $1.19\pm 0.08\gev$, which agrees with the observed 
statistical uncertainty of $1.15\gev$.

The same procedure with 5000 pseudo-datasets is applied to each of the seven top-quark mass values used 
for MC simulation, considering the correlation of the $R_{3/2}$ values for the
top quark and antiquark candidates in an event. Distributions of the pull values for the 5000 pseudo-datasets are derived,
where the pull is the difference between the fitted, $m_{t}^{\mathrm{fit}}$, and input,
$m_{t}^{\mathrm{inp}}$, top-quark mass values divided by the statistical uncertainty, $\sigma^{\mathrm{fit}}$, of the fit; 
$\mathrm{pull}=(m_{t}^{\mathrm{fit}}-m_{t}^{\mathrm{inp}})/\sigma^{\mathrm{fit}}$. 
The pull distribution for an unbiased measurement has a mean of zero and a standard deviation
of unity. For this measurement no dependence of the pull mean on 
$m_{t}^{\mathrm{inp}}$ is observed. An average pull mean value corresponding to 
$m_{t}^{\mathrm{fit}}-m_{t}^{\mathrm{inp}} = -0.23 \pm 0.14\gev$ 
and an average pull width of $1.175 \pm 0.027$ are obtained. 
The bias in the width of the pull is due to the statistical correlation.
To correct for this bias, the observed statistical uncertainty 
of $1.15\gev$ 
is scaled by 1.175 to yield the statistical uncertainty of $1.35\gev$ 
quoted in Eq.~(\ref{eqn-fitresult}). The 
bias indicated by the non-zero mean value of the pull distribution is corrected for 
in the above quoted result. The uncertainty of the pull mean value is considered as part 
of the systematic uncertainty related to the calibration of this measurement method.

\section{Systematic uncertainties}
\label{sec-systematics}
A large number of potential sources of systematic uncertainty were
evaluated. They can be categorised  as uncertainties due to: (i) the modelling of the
$\ttbar$ events in the MC simulation, (ii) the modelling of
the multijet background by the data-driven approach, (iii) the correction and
calibration of the energies of the reconstructed jets, the 
jet reconstruction and the $b$-quark identification efficiency. These are described in detail in 
Sects.~\ref{sec-SignalSystematics}--\ref{sec-JetSystematics}.
In general, for every investigated source of systematic 
uncertainty the likelihood fit of Eq.~(\ref{equ:bLike}) for the top-quark mass is 
repeated with a modified parameter. Any change of the measured top-quark mass 
is assigned as the systematic uncertainty due to this 
source. The total systematic uncertainty arises from adding all individual contributions 
in quadrature.
Table~\ref{tab:results} lists the individual contributions and
their combination.
The largest systematic uncertainties are due to the jet and $b$-jet energy scales and the ha\-dron\-isation modelling.

\subsection{Signal modelling}
\label{sec-SignalSystematics}
All systematic uncertainties related to the modelling of $\ttbar$ events and the lineshape
of the top-quark mass distribution are investigated using 5000 data sets, created by the resampling
technique described in Sect.~\ref{sec-TopMass} by randomly selecting $R_{3/2}$ values from a
distribution of $\ttbar$ MC 
simulation events generated with a shifted value for the relevant parameter as detailed below. In 
Table~\ref{tab:results}, the difference between the mean values obtained 
with shifted and with default parameter values, from 5000 pseudo-experiments each, 
is quoted for the investigated sources of systematic uncertainty.
\begin{table}%[tbp!]
\centerline{%\begin{center}
 \begin{tabular}{|l|r|}
\hline
\multicolumn{1}{|c|}{Signal modelling:}   & $\Delta m_{t}$ [\gev]   \\ \hline     
Method calibration                       & 0.42 \\
Trigger                                  & 0.01 \\
Signal MC generator                      & 0.30 \\
Hadronisation                            & 0.50 \\
Fast simulation                          & 0.24 \\
Colour reconnection                      & 0.22 \\
Underlying event                         & 0.08 \\
ISR and FSR                              & 0.22 \\
Proton PDF                               & 0.09 \\
Pile-up                                  & 0.02 \\ \hline \hline
\multicolumn{1}{|c|}{Background modelling:}& $\Delta m_{t}$ [\gev]   \\ \hline
Multijet background                      & 0.35 \\ \hline \hline
\multicolumn{1}{|c|}{Jet measurements:}   & $\Delta m_{t}$ [\gev]   \\ \hline
Jet energy scale (see Table~\ref{tab:jesresults}) & 0.51\\
$b$-jet energy scale                     & 0.62 \\ 
Jet energy resolution                    & 0.01 \\ 
Jet reconstruction efficiency            & 0.01 \\
$b$-tag efficiency and mistag rate       & 0.17 \\
Soft contributions to missing energy     & 0.02 \\ 
JVF scale factors                        & 0.02 \\ \hline\hline
Total systematic uncertainty             & 1.22 \\ \hline
\end{tabular}
}%
\caption{\label{tab:results}%
         Compilation of investigated systematic uncertainties on the determined top-quark mass reported
         in Sect.~\ref{sec-TopMass}. The three parts of the table correspond to uncertainties in 
         the $\ttbar$ and multijet background modelling, and uncertainties in the jet measurements.
}
\end{table}

\noindent{\bf Method calibration:} 
Our particular choice of signal parameterisation functions and the adopted linear dependence 
of the parameters of these functions on the top-quark mass value can affect the reconstructed 
top-quark mass. This uncertainty is estimated from the differences between the fitted and the 
input top-quark mass value when determining the $\ttbar$ template for each of the seven 
simulation samples separately. The average of the absolute differences is $0.23\gev$
and also accounts for the average shift of the pull distributions.

The shapes of the templates for $\ttbar$ and multijet background events can be
affected by statistical uncertainties of either simulated events (signal templates) or 
data (background templates). This is assessed by creating 1000 new sets of templates by
letting the standard templates fluctuate within their statistical uncertainties. The top-quark 
mass values obtained with these new templates are found to have an RMS spread
of $0.42\gev$.

The larger of $0.23\gev$ and $0.42\gev$ 
is assigned as a systematic uncertainty for the method calibration.

\noindent{\bf Trigger:}
Studies of the trigger efficiency close to the threshold region reveal a 5\% difference between data and MC 
simulation. The impact of this deviation is evaluated by reweighting the efficiency for
triggering MC simulation events to match the efficiency observed in data as a function of the transverse momentum of the fifth leading jet. The observed 
change in the measured top-quark mass is $0.01\gev$.

\noindent{\bf Signal MC generator:}
The impact of the choice of {\sc Powheg-box} as the signal MC generator 
is evaluated by generating $\ttbar$ events at $m_{t}=$ $172.5\gev$
using either {\sc Powheg-box} or {\sc MC@NLO}~\cite{FRI-0201,FRI-0301}, each
with {\sc Herwig}~\cite{COR-0001} for the modelling of the parton shower and the 
hadronisation. The full difference in the top-quark mass values of $0.30\gev$ found from 
using {\sc Powheg} or {\sc MC@NLO} to determine the signal templates is quoted as the systematic 
uncertainty.

\noindent{\bf Hadronisation:}
Potential systematic uncertainties due to our choice of parton shower and hadronisation model are assessed 
by using {\sc Powheg} $\ttbar$ events with parton shower and hadronisation
performed by either {\sc Pythia} with the {\sc Perugia P2012} tune or by {\sc Herwig}\footnote{%
Version 6.520 of {\sc Herwig} was
used with default parameters (expect for {\sc clpow}\ $=1.2$).
}
and {\sc Jimmy}
with the ATLAS AUET2 tune~\cite{ATL-PHYS-PUB-2011-008}. 
The full difference in the top-quark mass values of $0.50\gev$ between these two samples 
is ascribed to the uncertainty due to parton shower and hadronisation modelling.

\noindent{\bf Fast simulation:}
The $\ttbar$ MC simulation events for all seven $m_{t}$ mass values are processed by a 
fast simulation of the ATLAS detector~\cite{ATLAS:2010bfa,Edmonds:2008zz}. For $m_{t}=172.5\gev$ an additional 
$\ttbar$ MC simulation sample is created using the full simulation of the ATLAS detector. 
The systematic uncertainty of $0.24\gev$ is estimated from the difference of $0.24\pm0.30\stat\gev$ 
between the top-quark masses obtained by  
performing pseudo-experiments on either the fast or the full MC simulation sample.

\noindent{\bf Colour reconnection:}
Consequences of reconnection of colour flux lines between the partons 
are estimated with {\sc Powheg-box} and {\sc Pythia} by comparing simulated 
$\ttbar$ events based on the {\sc Perugia 2012} tune including colour reconnection (CR) 
and the {\sc Perugia 2012} {\sc loCR} tune~\cite{Skands:2010ak}, which uses a lower
colour reconnection strength than the default tune.
The full difference
of $0.22\gev$ in measured top-quark mass between these two samples is attributed to the 
uncertainty from colour reconnection.

\noindent{\bf Underlying event:}
The potential uncertainty due to the choice of a particular model to simulate underlying events
is evaluated by considering events simulated using {\sc Powheg-box} and 
{\sc Pythia} based on the {\sc Perugia} 2012 tune and comparing to events 
based on the {\sc Perugia 2012} {\sc mpiHi} tune~\cite{Skands:2010ak}, which has 
an increased rate of jets from multi-parton interactions.
Both tunes use
the same parameters for the modelling of colour reconnection and both predict 
similar activity in the plane transverse to the leading charged particle.
The samples used for colour reconnection uncertainties are based on 
different values for these parameters.
The full 
difference between the fitted mass values of $0.08\gev$ is taken as the systematic uncertainty.

\noindent{\bf Initial- and final-state QCD radiation:}
The impact from additional jets due to initial- and final-state QCD radiation, ISR and FSR,
respectively, on the
top-quark mass measurement is analysed with dedicated $\ttbar$ event samples generated
with the leading-order generator {\sc AcerMC}~\cite{Kersevan:2004yg}. Parton showering
and hadronisation are performed by {\sc Pythia} using the {\sc Perugia 2011C} tune.
Tunable parameters that control the parton shower strength are varied up and down in 
these samples in a range for which the simulated radiation in $\ttbar$ events is compatible 
with the results found from an investigation of additional jets in $\ttbar$ 
events~\cite{ATLAS:2012al}. Half of the full difference between the measured top-quark masses
from these two samples is taken as the systematic uncertainty, which is $0.22\gev$.

\noindent{\bf Proton Parton Distribution Function:}
The $\ttbar$ event samples were generated using CT10 PDF. The uncertainties in 
these PDFs are specified by 26 pairs of
additional PDF sets provided by the CTEQ group~\cite{cteq6l}. The effect of the
PDF uncertainties on the $\ttbar$ templates is derived from samples 
generated using {\sc MC@NLO} with {\sc Herwig} for hadronisation. 
For every additional PDF set, the simulated events are reweighted 
by the ratio of the varied PDF to the central PDF. Signal templates are
constructed for each of these 26 pairs of sets. Using these 
templates, pseudo-experiments are performed per pair of PDF sets but using the same
events for the up and down variations within every pair to alleviate the effects
of the statistical fluctuations. Half of the sum in quadrature of the difference
within each of the 26 pairs is assigned as the systematic uncertainty derived
from the CTEQ PDF.
Additionally, the $\ttbar$ event samples are also reweighted to the
central PDF set of either MSTW\-2008\cite{Martin:2009iq} or
NNPDF23\cite{Ball:2012cx}. The final systematic uncertainty due to PDF
is the sum in quadrature of these three contributions, which yields $0.09\gev$.

\noindent{\bf Pile-up:}
The consequences of additional proton--proton interactions 
on the top-quark mass measurement are investigated 
by repeating the full analysis separately
as a function of the number of reconstructed collision
vertices, $n_{\mathrm{vtx}}$, and as a function of the average 
number, $\langle\mu\rangle$, of inelastic 
proton--proton interactions per bunch crossing. 
This is in addition to the effects already accounted for in the 
corresponding jet energy scale. The data sample is split 
into disjoint subsamples of $n_{\mathrm{vtx}}\leq 5$, $5<n_{\mathrm{vtx}}\leq 7$, and $7<n_{\mathrm{vtx}}$,
or into subsamples of $\langle\mu\rangle\leq 6$,  
$6<\langle\mu\rangle\leq 10$, and $10<\langle\mu\rangle$.
In each of these subsamples the full analysis for the top-quark mass measurement
is repeated, giving per-subsample variations, $\Delta m_{t}$.
Within large statistical uncertainties, data and MC simulation agree. 
The effect of any residual differences between data and simulation is included 
by scaling $\Delta m_{t}$ with the absolute difference between the $n_{\mathrm{vtx}}$ 
distribution in data and simulation, each 
normalised to unit integral. The scaled $\Delta m_{t}$ obtained for each of the 
three subsamples are summed, yielding $0.02\gev$. %(?)
The same procedure is applied to the $\Delta m_{t}$ from the subsamples of the $\langle\mu\rangle$ 
distribution, yielding $0.01\gev$. %(?).
The two sums, derived from the $n_{\mathrm{vtx}}$ and for $\langle\mu\rangle$ distributions,
are then added in quadrature to estimate the systematic uncertainty on the top-quark mass measurement
of $0.02\gev$.

\subsection{Background modelling}
\label{sec-BackgroundSystematics}
Each of the prescriptions in Eq.~(\ref{eqn-ABEF-CDEF}) yields an independent estimate of 
the multijet background to the $\ttbar$ events. Employing these separately
distinguishes different contributions from
background processes and accounts for conceivable correlations 
between the distribution $N_F^{\mathrm{bkg}}(x)$ and the multiplicity of the $b$-tagged 
jets. In particular, the regions $C$ and $D$, where one jet is $b$-tagged, accumulate background 
from single top-quark production while suppressing contributions from $W$ $+$ jets processes. 
The regions $A$ and $B$, where no jets are $b$-tagged, are essentially free from $\ttbar$ 
events and, hence, insensitive to systematic uncertainties from the subtraction 
of residual $\ttbar$ contributions (see Eq.~(\ref{eqn-top_bkg_subtraction})). 
The average of the absolute shifts on $m_{t}$ when using either of the prescriptions in Eq.~(\ref{eqn-ABEF-CDEF}) 
separately is 
taken as symmetric uncertainty on the background modelling,
which amounts to $0.35\gev$.

\subsection{Jet measurement}
\label{sec-JetSystematics}
Systematic uncertainties due to measuring jets are listed in Table~\ref{tab:results} 
and detailed in the following.

\noindent{\bf Jet energy scale:}
The relative jet energy scale uncertainty varies between about $1\%$ and $3\%$ 
depending on the $\pT$ and $\eta$ of the jet. This was investigated in detail in 
Refs.~\cite{Aad:2014bia, Aad:2011he, ATLAS-CONF-2013-004}, which prescribe 21 components 
of uncertainty, including a proper treatment of the correlations between the 
individual sources. 
The 21 components involve nuisance parameters from different in situ techniques
applied to evaluate residual jet energy scale correction factors which account
for differences between data and MC simulation. They originate
from the calibration method, the calorimeter response, the detector simulation
and the specific choice of parameters in the physics model employed by the 
MC event generator. Further sources of uncertainty are related to
the extrapolation to the high-$\pT$ region, to the intercalibration of jets
at large pseudorapidity with central jets and to the pile-up. Topology-dependent
uncertainties arising from the relative numbers of jets initiated by gluons and light quarks 
are included as well as uncertainties on the response to jets with 
nearby hadronic activity. The 21 components
are considered uncorrelated. After repeating the top-quark mass measurement separately
for each component, the variation in the top-quark mass value obtained from the 
up and down variation of each nuisance parameter is symmetrised. The individual 
symmetrised contributions are added in quadrature to estimate the 
overall $\Delta m_{t}$ due to jet energy scale uncertainty of $0.51\gev$.

Table~\ref{tab:jesresults} lists the individual systematic uncertainty components
related to the energy measurements of jets combined into different categories 
according to the type of source and correlations (see Ref.\cite{Aad:2014bia}).

%%%%%%%%%%%%%%%%%%%%%%%%%%%%%%%%%%%%%%%%%%%%%%%%%%%%%%%%%%%%%%%%%%%%%%%%%%%%%%%
%% Table with total of subcomponents:
%%%%%%%%%%%%%%%%%%%%%%%%%%%%%%%%%%%%%%%%%%%%%%%%%%%%%%%%%%%%%%%%%%%%%%%%%%%%%%%
\begin{table}%[tbp!]
\begin{center}
\begin{tabular}{|l||r|}
\cline{2-2}
\multicolumn{1}{c|}{}                             & $\Delta m_{t}$ [\gev] \\ \hline
\textrm{Statistics and method}                    &  \textrm{0.09} \\
\textrm{Physics modelling}                        &  \textrm{0.31} \\
\textrm{Detector description}                     &  \textrm{0.36}\\
\textrm{Mixed detector and modelling}             &  \textrm{0.05}\\
\textrm{Single high-$\pT$ particle}               &  \textrm{0.02}\\
\textrm{Relative non-closure in MC}               &  \textrm{0.04} \\
\textrm{Pile-up}                                  &  \textrm{0.03} \\
\textrm{Close-by jets}                            &  \textrm{0.02} \\
\textrm{Flavour composition and response}         &  \textrm{0.10}\\
\hline \hline
\textrm{Jet energy scale}                         &  \textrm{0.51} \\
\hline
\textrm{$b$-jet energy scale }                    &  \textrm{0.62} \\
\hline
\end{tabular}
\end{center}
\caption{\label{tab:jesresults}%
         Individual contributions to the systematic uncertainty of the top-quark mass 
         due to uncertainties on the jet energy scale listed in Table~\ref{tab:results}.
}
\end{table}
%%%%%%%%%%%%%%%%%%%%%%%%%%%%%%%%%%%%%%%%%%%%%%%%%%%%%%%%%%%%%%%%%%%%%%%%%%%%%%%

\noindent{\bf Relative $b$-jet energy scale:}
The relative $b$-jet energy scale accounts for the remaining differences between 
an inclusive jets sample and jets originating from bottom quarks after the 
global jet energy scale is determined. It is estimated by choosing different 
fragmentation models.
An extra uncertainty, ranging between $1.8\%$ and $0.7\%$, and decreasing as jet $\pT$ increases,
is assigned to each $b$-jet to account for the difference 
between jets containing $b$-flavoured hadrons and the inclusive jet sample. This
uncertainty is derived from MC simulation studies and validated 
by comparison with data (see Ref.~\cite{ATLAS-CONF-2013-002} for details). For the 
spectrum of jets selected in this analysis the average uncertainty is less
than $1.2\%$. The systematic uncertainty on $m_t$ due to the relative $b$-jet energy
scale is $0.62\gev$.

%\subsection{Jet measurement}
%\label{sec-JetSystematics}
%
\noindent{\bf Jet energy resolution:}
The impact of a residual difference between the jet energy resolution in
data and MC simulation is accounted for by smearing
the energy of each reconstructed jet in the simulation by a Gaussian
function before applying the event selection requirements (see Ref.~\cite{Aad:2012ag}
for details). 
The top-quark mass measurement is repeated using the smeared
jet energies yielding a variation of $0.01\gev$, which is symmetrised and assigned as a systematic uncertainty.

\noindent{\bf Jet reconstruction efficiency:}
The jet reconstruction efficiency was found in Ref.~\cite{Aad:2011he} to
differ in data and MC simulation by no more than $\pm2\%$. 
This residual difference is applied as a variation by randomly removing jets 
from the simulated events before applying the event selection criteria. 
The variation of $0.01\gev$ 
found by repeating the top-quark mass measurement employing this modified MC 
simulation sample is taken as a systematic uncertainty.

\noindent{\bf $b$-tagging efficiency and mistag rate:}
The efficiency for tagging $b$-quark jets as well as the $c$-quark and light-quark
($u$, $d$, $s$) jet mistag rate in simulation are corrected to data by scale 
factors~\cite{ATLAS-CONF-2012-040, ATLAS-CONF-2012-043}. 
The uncertainty of this correction is propagated to the measured top-quark 
mass by varying these scale factors by one standard deviation about their
central values, which 
depend on the
$\pT$ and the $\eta$ of the jet, and on the underlying quark flavour. 
The variations
in the top-quark mass 
are added in quadrature to assess the systematic uncertainty from
this source, which yields $0.17\gev$.

\noindent{\bf Soft contribution to missing energy:}
Measured energy deposits in the calorimeter which are not associated with 
a high-$\pT$ jet, photon, electron, or muon, stem mostly from low-$\pT$ 
particles. 
These energy deposits are calibrated 
using the local ha\-dron\-ic calibration scheme\cite{Barillari:2009zza}.
An uncertainty of $0.02\gev$ on the top-quark mass
due to this assumption 
is derived by scaling the soft contributions within their uncertainties.

\noindent{\bf Jet vertex fraction scale factor uncertainty:}
The difference in JVF between data and MC 
simulation is corrected by applying scale factors. These scale factors,
varied according to their uncertainty, are applied to MC
simulation events as a function of the $\pT$ of a jet. The resulting variation in the measured 
top-quark mass 
amounts to $0.02\gev$. 

\section{Comparison with alternative analysis}
\label{sec-Validation}
The result of this measurement is compared with an independent 
measurement based on essentially the same selection described in Sect.~\ref{sec-DatSimSel}.
For this independent measurement, however, entirely different methods are chosen for alleviating 
the effects due to 
uncertainties from the jet energy measurement
and for modelling the 
multijet background. Applying a simultaneous two-di\-men\-sion\-al fit to the $W$ boson and
top-quark masses unfolds the dependency
of the top-quark mass on a global jet scale factor. Thus systematic uncertainties
affecting the jet scale factor are mostly removed from the uncertainties in the measured
top-quark mass; however, this gives rise to increased statistical uncertainty (see also Ref.~\cite{ATLAS:2012aj}). 

In the independent alternative measurement, the multijet background is modelled using 
an event mixing procedure. Here, events with six or
more jets are composed from events with exactly five jets, two of which are $b$-tagged, merged with the sixth and
subsequent leading jets from events of an independent inclusive jet sample.
Kinematic similarity of the two events to be mixed is ensured by requiring the
similarity of the transverse momenta of both the leading jets in the two events 
and also of the fifth leading jets. Evaluation of the systematic uncertainties described in
Sect.~\ref{sec-systematics} was performed for this independent 
analysis. This investigation showed that 
the alternative analysis and the main analysis have similar sensitivities to the top-quark mass.
The alternative analysis has yielded a top-quark mass value and a total statistical uncertainty of $m_{t} = 174.7 \pm 1.4\ (\mathrm{stat.+JSF}) \gev$ 
with a global jet scale factor of % $\mathrm{JSF} = 1.0128 \pm 0.0075\, (\mathrm{stat.})$,
                                    $\mathrm{JSF} = 1.013 \pm 0.008\, (\mathrm{stat.})$,
%and systematic uncertainty 
in good agreement with the results presented
in Sects.~\ref{sec-TopMass} and \ref{sec-systematics}.

\section{Summary}
In a data set corresponding to $4.6\,\ifb$ of proton--proton collisions
collected by the ATLAS experiment at the LHC at $\sqrt{s}=7\tev$,
events consistent with $\ttbar$ pairs decaying into a fully hadronic final state were selected.
 A kinematic likelihood fit was employed to assign reconstructed jets to the partons expected from the 
leading-order
hadronic decay of the intermediate $\ttbar$ state. To reduce the sensitivity of the analysis
to the energy scale of jets, the ratio $R_{3/2}$ of the three-jet mass to the 
dijet mass was calculated. The three-jet mass calculation combines all jets from a top-quark 
decay, and the dijet mass is computed with the two jets from the hadronically decaying $W$ boson.
The multijet background was 
determined by dividing the event sample into six disjoint sets according to the
number of $b$-tagged jets and the $\pT$ of the sixth jet. 
The background in the region of interest is then estimated by cross-multiplication.
Fitting the $R_{3/2}$ distribution for the top-quark mass yields 
\begin{equation}
  m_{t} = 175.1 \pm 1.4\stat \pm 1.2\syst \gev 
\end{equation}
with a measured fraction of background events 
 $f_{\mathrm{bkg}}=0.72\pm 0.01$.
The systematic uncertainties are dominated by the residual uncertainties from the
jet energy scale for all jets and, specifically, for $b$-quark jets and by the uncertainties from
hadronisation modelling. 
The total uncertainty is $1.8\gev$. 
This result has a precision similar to, and within uncertainties fully agrees with, the top-quark mass measured from the fully hadronic final state 
by other experiments~\cite{Chatrchyan:2013xza,Aaltonen:2011em} and the result measured in the lepton plus jets final state and published previously by ATLAS~\cite{ATLAS:2012aj}.

%%%%%%%%%%%%%%%
% Acknowledgements
%%%%%%%%%%%%%%%
% Acknowledgements for papers with collision data
% Version 19-Feb-2015

\section{Acknowledgements}

% Standard acknowledgements start here
%----------------------------------------------
We thank CERN for the very successful operation of the LHC, as well as the
support staff from our institutions without whom ATLAS could not be
operated efficiently.

We acknowledge the support of ANPCyT, Argentina; YerPhI, Armenia; ARC,
Australia; BMWFW and FWF, Austria; ANAS, Azerbaijan; SSTC, Belarus; CNPq and FAPESP,
Brazil; NSERC, NRC and CFI, Canada; CERN; CONICYT, Chile; CAS, MOST and NSFC,
China; COLCIENCIAS, Colombia; MSMT CR, MPO CR and VSC CR, Czech Republic;
DNRF, DNSRC and Lundbeck Foundation, Denmark; EPLANET, ERC and NSRF, European Union;
IN2P3-CNRS, CEA-DSM/IRFU, France; GNSF, Georgia; BMBF, DFG, HGF, MPG and AvH
Foundation, Germany; GSRT and NSRF, Greece; RGC, Hong Kong SAR, China; ISF, MINERVA, GIF, I-CORE and Benoziyo Center, Israel; INFN, Italy; MEXT and JSPS, Japan; CNRST, Morocco; FOM and NWO, Netherlands; BRF and RCN, Norway; MNiSW and NCN, Poland; GRICES and FCT, Portugal; MNE/IFA, Romania; MES of Russia and ROSATOM, Russian Federation; JINR; MSTD,
Serbia; MSSR, Slovakia; ARRS and MIZ\v{S}, Slovenia; DST/NRF, South Africa;
MINECO, Spain; SRC and Wallenberg Foundation, Sweden; SER, SNSF and Cantons of
Bern and Geneva, Switzerland; NSC, Taiwan; TAEK, Turkey; STFC, the Royal
Society and Leverhulme Trust, United Kingdom; DOE and NSF, United States of
America.

The crucial computing support from all WLCG partners is acknowledged
gratefully, in particular from CERN and the ATLAS Tier-1 facilities at
TRIUMF (Canada), NDGF (Denmark, Norway, Sweden), CC-IN2P3 (France),
KIT/GridKA (Germany), INFN-CNAF (Italy), NL-T1 (Netherlands), PIC (Spain),
ASGC (Taiwan), RAL (UK) and BNL (USA) and in the Tier-2 facilities
worldwide.
%----------------------------------------------

%\clearpage
%%%%%%%%%%%%%%%
% Bibliography
%%%%%%%%%%%%%%%
\bibliographystyle{atlas4epjc}
\bibliography{mtoppap-epjc}

\providecommand{\href}[2]{#2}\begingroup\raggedright\begin{thebibliography}{10}

\bibitem{bib-HiggsDiscoveryPaper-ATLAS}
{ATLAS} Collaboration, { {Observation of a new particle in the search for the
  Standard Model Higgs boson with the ATLAS detector at the LHC}\/},
  \href{http://dx.doi.org/10.1016/j.physletb.2012.08.020}{Phys.\ Lett.\ B {\bf
  716}, 1--29 (2012)},
\href{http://arxiv.org/abs/1207.7214}{{\tt arXiv:1207.7214 [hep-ex]}}.
%%CITATION = ARXIV:1207.7214;%%.

\bibitem{bib-HiggsDiscoveryPaper-CMS}
{CMS} Collaboration, { {Observation of a new boson at a mass of $125\gev$ with
  the CMS experiment at the LHC}\/},
  \href{http://dx.doi.org/10.1016/j.physletb.2012.08.021}{Phys.\ Lett.\ B {\bf
  716}, 30--61 (2012)},
\href{http://arxiv.org/abs/1207.7235}{{\tt arXiv:1207.7235 [hep-ex]}}.
%%CITATION = ARXIV:1207.7235;%%.

\bibitem{Sher:1988mj}
M.~Sher, { {Electroweak Higgs Potentials and Vacuum Stability}\/},
\href{http://dx.doi.org/10.1016/0370-1573(89)90061-6}{Phys.\ Rept. {\bf 179},
  273--418 (1989)}.
%%CITATION = PRPLC,179,273;%%.

\bibitem{Baak:2012kk}
M.~Baak et al., { {The Electroweak Fit of the Standard Model after the
  Discovery of a New Boson at the LHC}\/},
  \href{http://dx.doi.org/10.1140/epjc/s10052-012-2205-9}{Eur.\ Phys.\ J.\ C
  {\bf 72}, 2205 (2012)},
\href{http://arxiv.org/abs/1209.2716}{{\tt arXiv:1209.2716 [hep-ph]}}.
%%CITATION = ARXIV:1209.2716;%%.

\bibitem{ATLAS:2014wva}
{ATLAS Collaboration, CDF Collaboration, CMS Collaboration, D0} Collaboration,
  { {First combination of Tevatron and LHC measurements of the top-quark
  mass}\/},  ATLAS-CONF-2014-008, CDF-NOTE-11071, CMS-PAS-TOP-13-014,
  D0-NOTE-6416, \href{http://arxiv.org/abs/1403.4427}{{\tt arXiv:1403.4427
  [hep-ex]}}.
\url{http://inspirehep.net/record/1286320}.
%%CITATION = ARXIV:1403.4427;%%.

\bibitem{Abazov:2014dpa}
{D0} Collaboration, V.~M. Abazov et al., { {Precision measurement of the
  top-quark mass in lepton+jets final states}\/},
  \href{http://dx.doi.org/10.1103/PhysRevLett.113.032002}{Phys.\ Rev.\ Lett.
  {\bf 113}, 032002 (2014)},
\href{http://arxiv.org/abs/1405.1756}{{\tt arXiv:1405.1756 [hep-ex]}}.
%%CITATION = ARXIV:1405.1756;%%.

\bibitem{Abazov:2012rp}
{D0} Collaboration, V.~M. Abazov et al., { {Measurement of the top quark mass
  in $p \bar{p}$ collisions using events with two leptons}\/},
  \href{http://dx.doi.org/10.1103/PhysRevD.86.051103}{Phys.\ Rev.\ D {\bf 86},
  051103 (2012)},
\href{http://arxiv.org/abs/1201.5172}{{\tt arXiv:1201.5172 [hep-ex]}}.
%%CITATION = ARXIV:1201.5172;%%.

\bibitem{Aaltonen:2012va}
{CDF} Collaboration, T.~Aaltonen et al., { {Precision Top-Quark Mass
  Measurements at CDF}\/},
  \href{http://dx.doi.org/10.1103/PhysRevLett.109.152003}{Phys.\ Rev.\ Lett.
  {\bf 109}, 152003 (2012)},
\href{http://arxiv.org/abs/1207.6758}{{\tt arXiv:1207.6758 [hep-ex]}}.
%%CITATION = ARXIV:1207.6758;%%.

\bibitem{Aaltonen:2011dr}
{CDF} Collaboration, T.~Aaltonen et al., { {Top quark mass measurement using
  the template method at CDF}\/},
  \href{http://dx.doi.org/10.1103/PhysRevD.83.111101}{Phys.\ Rev.\ D {\bf 83},
  111101 (2011)},
\href{http://arxiv.org/abs/1105.0192}{{\tt arXiv:1105.0192 [hep-ex]}}.
%%CITATION = ARXIV:1105.0192;%%.

\bibitem{Aaltonen:2011em}
{CDF} Collaboration, T.~Aaltonen et al., { {Measurement of the Top Quark Mass
  in the All-Hadronic Mode at CDF}\/},
  \href{http://dx.doi.org/10.1016/j.physletb.2012.06.007}{Phys.\ Lett.\ B {\bf
  714}, 24--31 (2012)},
\href{http://arxiv.org/abs/1112.4891}{{\tt arXiv:1112.4891 [hep-ex]}}.
%%CITATION = ARXIV:1112.4891;%%.

\bibitem{Chatrchyan:2013xza}
{CMS} Collaboration, { {Measurement of the top-quark mass in all-jets $\ttbar$
  events in pp collisions at $\sqrt{s}=7\tev$}\/},
  \href{http://dx.doi.org/10.1140/epjc/s10052-014-2758-x}{Eur.\ Phys.\ J.\ C
  {\bf 74}, 2758 (2014)},
\href{http://arxiv.org/abs/1307.4617}{{\tt arXiv:1307.4617 [hep-ex]}}.
%%CITATION = ARXIV:1307.4617;%%.

\bibitem{Chatrchyan:2013boa}
{CMS} Collaboration, { {Measurement of masses in the $t \bar{t}$ system by
  kinematic endpoints in pp collisions at $\sqrt{s} = 7\tev$}\/},
  \href{http://dx.doi.org/10.1140/epjc/s10052-013-2494-7}{Eur.\ Phys.\ J.\ C
  {\bf 73}, 2494 (2013)},
\href{http://arxiv.org/abs/1304.5783}{{\tt arXiv:1304.5783 [hep-ex]}}.
%%CITATION = ARXIV:1304.5783;%%.

\bibitem{Chatrchyan:2012ea}
{CMS} Collaboration, { {Measurement of the top-quark mass in $t\bar{t}$ events
  with dilepton final states in $pp$ collisions at $\sqrt{s}=7\tev$}\/},
  \href{http://dx.doi.org/10.1140/epjc/s10052-012-2202-z}{Eur.\ Phys.\ J.\ C
  {\bf 72}, 2202 (2012)},
\href{http://arxiv.org/abs/1209.2393}{{\tt arXiv:1209.2393 [hep-ex]}}.
%%CITATION = ARXIV:1209.2393;%%.

\bibitem{Chatrchyan:2012cz}
{CMS} Collaboration, { {Measurement of the top-quark mass in $t\bar{t}$ events
  with lepton+jets final states in $pp$ collisions at $\sqrt{s}=7\tev$}\/},
  \href{http://dx.doi.org/10.1007/JHEP12(2012)105}{JHEP {\bf 1212}, 105
  (2012)},
\href{http://arxiv.org/abs/1209.2319}{{\tt arXiv:1209.2319 [hep-ex]}}.
%%CITATION = ARXIV:1209.2319;%%.

\bibitem{ATLAS:2012aj}
{ATLAS} Collaboration, { {Measurement of the top quark mass with the template
  method in the $t \bar{t}\rightarrow $ lepton + jets channel using ATLAS
  data}\/},  \href{http://dx.doi.org/10.1140/epjc/s10052-012-2046-6}{Eur.\
  Phys.\ J.\ C {\bf 72}, 2046 (2012)},
\href{http://arxiv.org/abs/1203.5755}{{\tt arXiv:1203.5755 [hep-ex]}}.
%%CITATION = ARXIV:1203.5755;%%.

\bibitem{Aad:2008zzm}
{ATLAS} Collaboration, { {The ATLAS Experiment at the CERN Large Hadron
  Collider}\/},
\href{http://dx.doi.org/10.1088/1748-0221/3/08/S08003}{JINST {\bf 3}, S08003
  (2008)}.
%%CITATION = JINST,3,S08003;%%.

\bibitem{Aad:2011he}
{ATLAS} Collaboration, { {Jet energy measurement with the ATLAS detector in
  proton-proton collisions at $\sqrt{s}=7\tev$}\/},
  \href{http://dx.doi.org/10.1140/epjc/s10052-013-2304-2}{Eur.\ Phys.\ J.\ C
  {\bf 73}, 2304 (2013)},
\href{http://arxiv.org/abs/1112.6426}{{\tt arXiv:1112.6426 [hep-ex]}}.
%%CITATION = ARXIV:1112.6426;%%.

\bibitem{ATLAS-CONF-2011-102}
{ATLAS} Collaboration, { {Commissioning of the ATLAS high-performance b-tagging
  algorithms in the $7\tev$ collision data}\/},  ATLAS-CONF-2011-102.
  \url{http://inspirehep.net/record/1204180}.

\bibitem{ATLAS-CONF-2012-040}
{{ATLAS}} Collaboration, { Measurement of the Mistag Rate of b-tagging
  algorithms with $5\ifb$ of Data Collected by the ATLAS Detector\/},
  ATLAS-CONF-2012-040. \url{http://inspirehep.net/record/1204279}.

\bibitem{ATLAS-CONF-2012-097}
{ATLAS} Collaboration, { {Measuring the $b$-tag efficiency in a top-pair sample
  with $4.7\ifb$ of data from the ATLAS detector}\/},  ATLAS-CONF-2012-097.
  \url{http://inspirehep.net/record/1204320}.

\bibitem{Aad:2013ucp}
{ATLAS} Collaboration, { {Improved luminosity determination in $pp$ collisions
  at $\sqrt{s} = 7\tev$ using the ATLAS detector at the LHC}\/},
  \href{http://dx.doi.org/10.1140/epjc/s10052-013-2518-3}{Eur.\ Phys.\ J.\ C
  {\bf 73}, 2518 (2013)},
\href{http://arxiv.org/abs/1302.4393}{{\tt arXiv:1302.4393 [hep-ex]}}.
%%CITATION = ARXIV:1302.4393;%%.

\bibitem{ATLAS:2010bfa}
{ATLAS} Collaboration, { {The simulation principle and performance of the ATLAS
  fast calorimeter simulation FastCaloSim}\/},  ATL-PHYS-PUB-2010-013.
\url{http://inspirehep.net/record/1194623}.
%%CITATION = ATL-PHYS-PUB-2010-013 ETC.;%%.

\bibitem{Agostinelli:2002hh}
{GEANT4} Collaboration, S.~Agostinelli et al., { {GEANT4: A Simulation
  toolkit}\/},
\href{http://dx.doi.org/10.1016/S0168-9002(03)01368-8}{Nucl.\ Instrum.\ Meth.\
  A {\bf 506}, 250--303 (2003)}.
%%CITATION = NUIMA,A506,250;%%.

\bibitem{Aad:2010ah}
{ATLAS} Collaboration, { {The ATLAS Simulation Infrastructure}\/},
  \href{http://dx.doi.org/10.1140/epjc/s10052-010-1429-9}{Eur.\ Phys.\ J.\ C
  {\bf 70}, 823--874 (2010)},
\href{http://arxiv.org/abs/1005.4568}{{\tt arXiv:1005.4568 [physics.ins-det]}}.
%%CITATION = ARXIV:1005.4568;%%.

\bibitem{Alioli:2010xd}
S.~Alioli, P.~Nason, C.~Oleari, and E.~Re, { {A general framework for
  implementing NLO calculations in shower Monte Carlo programs: the POWHEG
  BOX}\/},  \href{http://dx.doi.org/10.1007/JHEP06(2010)043}{JHEP {\bf 1006},
  043 (2010)}, \href{http://arxiv.org/abs/1002.2581}{{\tt arXiv:1002.2581
  [hep-ph]}}.

\bibitem{Frixione:2007nw}
S.~Frixione, P.~Nason, and G.~Ridolfi, { {A Positive-weight
  next-to-leading-order Monte Carlo for heavy flavour hadroproduction}\/},
  \href{http://dx.doi.org/10.1088/1126-6708/2007/09/126}{JHEP {\bf 0709}, 126
  (2007)},
\href{http://arxiv.org/abs/0707.3088}{{\tt arXiv:0707.3088 [hep-ph]}}.
%%CITATION = ARXIV:0707.3088;%%.

\bibitem{Lai:2010vv}
H.-L. Lai et al., { {New parton distributions for collider physics}\/},
  \href{http://dx.doi.org/10.1103/PhysRevD.82.074024}{Phys.\ Rev.\ D {\bf 82},
  074024 (2010)},
\href{http://arxiv.org/abs/1007.2241}{{\tt arXiv:1007.2241 [hep-ph]}}.
%%CITATION = ARXIV:1007.2241;%%.

\bibitem{Pythia}
T.~Sjostrand, S.~Mrenna, and P.~Z. Skands, { {PYTHIA 6.4 Physics and
  Manual}\/},  \href{http://dx.doi.org/10.1088/1126-6708/2006/05/026}{JHEP {\bf
  05}, 026 (2006)}, \href{http://arxiv.org/abs/hep-ph/0603175}{{\tt
  arXiv:hep-ph/0603175}}.
Version 6.425 was used.
%%CITATION = HEP-PH/0603175;%%.

\bibitem{Skands:2010ak}
P.~Z. Skands, { {Tuning Monte Carlo Generators: The Perugia Tunes}\/},
  \href{http://dx.doi.org/10.1103/PhysRevD.82.074018}{Phys.\ Rev.\ D {\bf 82},
  074018 (2010)},
\href{http://arxiv.org/abs/1005.3457}{{\tt arXiv:1005.3457 [hep-ph]}}.
%%CITATION = ARXIV:1005.3457;%%.

\bibitem{Aad:2012xs}
{ATLAS} Collaboration, { {Performance of the ATLAS Trigger System in 2010}\/},
  \href{http://dx.doi.org/10.1140/epjc/s10052-011-1849-1}{Eur.\ Phys.\ J.\ C
  {\bf 72}, 1849 (2012)},
\href{http://arxiv.org/abs/1110.1530}{{\tt arXiv:1110.1530 [hep-ex]}}.
%%CITATION = ARXIV:1110.1530;%%.

\bibitem{Cacciari:2008gp}
M.~Cacciari, G.~P. Salam, and G.~Soyez, { {The anti-$k_t$ jet clustering
  algorithm}\/},  \href{http://dx.doi.org/10.1088/1126-6708/2008/04/063}{JHEP
  {\bf 04}, 063 (2008)},
\href{http://arxiv.org/abs/0802.1189}{{\tt arXiv:0802.1189 [hep-ph]}}.
%%CITATION = 0802.1189;%%.

\bibitem{Lampl:1099735}
W.~Lampl et al., { {Calorimeter Clustering Algorithms: Description and
  Performance}\/},  ATL-LARG-PUB-2008-002.
  \url{http://inspirehep.net/record/807147}.

\bibitem{ATLAS-CONF-2012-20}
{{ATLAS}} Collaboration, { Selection of jets produced in proton-proton
  collisions with the ATLAS detector using 2011 data\/},  ATLAS-CONF-2012-020.
  \url{http://inspirehep.net/record/1204199}.

\bibitem{Aad:2014bia}
{{ATLAS}} Collaboration, { {Jet energy measurement and its systematic
  uncertainty in proton-proton collisions at $\sqrt{s}=7$ TeV with the ATLAS
  detector}\/},
  \href{http://dx.doi.org/10.1140/epjc/s10052-014-3190-y}{Eur.Phys.J. {\bf
  C75}, 17 (2015)},
\href{http://arxiv.org/abs/1406.0076}{{\tt arXiv:1406.0076 [hep-ex]}}.
%%CITATION = ARXIV:1406.0076;%%.

\bibitem{ATLAS-CONF-2013-004}
{{ATLAS}} Collaboration, { {Jet energy scale and its systematic uncertainty in
  proton-proton collisions at sqrt(s)=7 TeV with ATLAS 2011 data}\/},
  ATLAS-CONF-2013-004.
\url{http://inspirehep.net/record/1229978}.
%%CITATION = ATLAS-CONF-2013-004 ETC.;%%.

\bibitem{ATLAS-CONF-2013-002}
{{ATLAS}} Collaboration, { Jet energy measurement and systematic uncertainties
  using tracks for jets and for b-quark jets produced in proton-proton
  collisions at $\sqrt{s} = 7\tev$ in the ATLAS detector\/},
  ATLAS-CONF-2013-002. \url{http://inspirehep.net/record/1229980}.

\bibitem{CERN-PH-EP-2011-117}
{ATLAS} Collaboration, { {Electron performance measurements with the ATLAS
  detector using the 2010 LHC proton-proton collision data}\/},
  \href{http://dx.doi.org/10.1140/epjc/s10052-012-1909-1}{Eur.\ Phys.\ J.\ C
  {\bf 72}, 1909 (2012)},
\href{http://arxiv.org/abs/1110.3174}{{\tt arXiv:1110.3174 [hep-ex]}}.
%%CITATION = ARXIV:1110.3174;%%.

\bibitem{Aad:2014zya}
{ATLAS} Collaboration, { {Muon reconstruction efficiency and momentum
  resolution of the ATLAS experiment in proton-proton collisions at
  $\sqrt{s}$=7 TeV in 2010}\/},
  \href{http://dx.doi.org/10.1140/epjc/s10052-014-3034-9}{Eur.\ Phys.\ J.\ C
  {\bf 74}, 3034 (2014)},
\href{http://arxiv.org/abs/1404.4562}{{\tt arXiv:1404.4562 [hep-ex]}}.
%%CITATION = ARXIV:1404.4562;%%.

\bibitem{ATLAS-CONF-2012-043}
{ATLAS} Collaboration, { {Measurement of the $b$-tag Efficiency in a Sample of
  Jets Containing Muons with $5\ifb$ of Data from the ATLAS Detector}\/},
  ATLAS-CONF-2012-043.
\url{http://inspirehep.net/record/1204276}.
%%CITATION = ATLAS-CONF-2012-043 ETC.;%%.

\bibitem{Barillari:2009zza}
{ATLAS} Collaboration, { {Local hadronic calibration}\/},
  ATL-LARG-PUB-2009-001-2.
\url{http://inspirehep.net/record/811642}.
%%CITATION = ATL-LARG-PUB-2009-001-2 ETC.;%%.

\bibitem{Erdmann:2013rxa}
J.~Erdmann et al., { {A likelihood-based reconstruction algorithm for top-quark
  pairs and the KLFitter framework}\/},
  \href{http://dx.doi.org/10.1016/j.nima.2014.02.029}{Nucl.\ Instrum.\ Meth.\ A
  {\bf 748}, 18--25 (2014)},
\href{http://arxiv.org/abs/1312.5595}{{\tt arXiv:1312.5595 [hep-ex]}}.
%%CITATION = ARXIV:1312.5595;%%.

\bibitem{PDG}
{Particle Data Group} Collaboration, J.~Beringer et al., { {Review of Particle
  Physics (RPP)}\/},
\href{http://dx.doi.org/10.1103/PhysRevD.86.010001}{Phys.\ Rev.\ D {\bf 86},
  010001 (2012)}.
%%CITATION = PHRVA,D86,010001;%%.

\bibitem{MAN-HEP-99-4}
R.~Barlow, { {Application of the bootstrap resampling technique to particle
  physics experiments}\/},  MAN/HEP/99/4.
  \url{https://www.hep.manchester.ac.uk/preprints/manhep99-4.ps}.

\bibitem{FRI-0201}
S.~Frixione and B.~R. Webber, { {Matching NLO QCD computations and parton
  shower simulations}\/},
  \href{http://dx.doi.org/10.1088/1126-6708/2002/06/029}{JHEP {\bf 0206}, 029
  (2002)},
\href{http://arxiv.org/abs/hep-ph/0204244}{{\tt arXiv:hep-ph/0204244
  [hep-ph]}}.
%%CITATION = HEP-PH/0204244;%%.

\bibitem{FRI-0301}
S.~Frixione, P.~Nason, and B.~R. Webber, { {Matching NLO QCD and parton showers
  in heavy flavor production}\/},
  \href{http://dx.doi.org/10.1088/1126-6708/2003/08/007}{JHEP {\bf 0308}, 007
  (2003)},
\href{http://arxiv.org/abs/hep-ph/0305252}{{\tt arXiv:hep-ph/0305252
  [hep-ph]}}.
%%CITATION = HEP-PH/0305252;%%.

\bibitem{COR-0001}
G.~Corcella et al., { {HERWIG 6: An Event generator for hadron emission
  reactions with interfering gluons (including supersymmetric processes)}\/},
  \href{http://dx.doi.org/10.1088/1126-6708/2001/01/010}{JHEP {\bf 0101}, 010
  (2001)}, \href{http://arxiv.org/abs/hep-ph/0011363}{{\tt arXiv:hep-ph/0011363
  [hep-ph]}}.
Version 6.520 has been used.
%%CITATION = HEP-PH/0011363;%%.

\bibitem{ATL-PHYS-PUB-2011-008}
{ATLAS} Collaboration, { {New ATLAS event generator tunes to 2010 data}\/},
  ATL-PHYS-PUB-2011-008. \url{http://inspirehep.net/record/1196773}.

\bibitem{Edmonds:2008zz}
K.~Edmonds et al., { {The fast ATLAS track simulation (FATRAS)}\/},
  ATL-SOFT-PUB-2008-001.
\url{http://inspirehep.net/record/807206}.
%%CITATION = ATL-SOFT-PUB-2008-001 ETC.;%%.

\bibitem{Kersevan:2004yg}
B.~P. Kersevan and E.~Richter-Was, { {The Monte Carlo event generator AcerMC
  versions 2.0 to 3.8 with interfaces to PYTHIA 6.4, HERWIG 6.5 and ARIADNE
  4.1}\/},  \href{http://dx.doi.org/10.1016/j.cpc.2012.10.032}{Comput.\ Phys.\
  Commun. {\bf 184}, 919--985 (2013)},
\href{http://arxiv.org/abs/hep-ph/0405247}{{\tt arXiv:hep-ph/0405247
  [hep-ph]}}.
%%CITATION = HEP-PH/0405247;%%.

\bibitem{ATLAS:2012al}
{ATLAS} Collaboration, { {Measurement of $\ttbar$ production with a veto on
  additional central jet activity in pp collisions at $\sqrt{s} = 7\tev$ using
  the ATLAS detector}\/},
  \href{http://dx.doi.org/10.1140/epjc/s10052-012-2043-9}{Eur.\ Phys.\ J.\ C
  {\bf 72}, 2043 (2012)},
\href{http://arxiv.org/abs/1203.5015}{{\tt arXiv:1203.5015 [hep-ex]}}.
%%CITATION = ARXIV:1203.5015;%%.

\bibitem{cteq6l}
J.~Pumplin et al., { {New generation of parton distributions with uncertainties
  from global QCD analysis}\/},
  \href{http://dx.doi.org/10.1088/1126-6708/2002/07/012}{JHEP {\bf 0207}, 012
  (2002)},
\href{http://arxiv.org/abs/hep-ph/0201195}{{\tt arXiv:hep-ph/0201195
  [hep-ph]}}.
%%CITATION = HEP-PH/0201195;%%.

\bibitem{Martin:2009iq}
A.~Martin, W.~Stirling, R.~Thorne, and G.~Watt, { {Parton distributions for the
  LHC}\/},  \href{http://dx.doi.org/10.1140/epjc/s10052-009-1072-5}{Eur.\
  Phys.\ J.\ C {\bf 63}, 189--285 (2009)},
\href{http://arxiv.org/abs/0901.0002}{{\tt arXiv:0901.0002 [hep-ph]}}.
%%CITATION = ARXIV:0901.0002;%%.

\bibitem{Ball:2012cx}
R.~D. Ball et al., { {Parton distributions with LHC data}\/},
  \href{http://dx.doi.org/10.1016/j.nuclphysb.2012.10.003}{Nucl.\ Phys.\ B {\bf
  867}, 244--289 (2013)},
\href{http://arxiv.org/abs/1207.1303}{{\tt arXiv:1207.1303 [hep-ph]}}.
%%CITATION = ARXIV:1207.1303;%%.

\bibitem{Aad:2012ag}
{ATLAS} Collaboration, { {Jet energy resolution in proton-proton collisions at
  $\sqrt{s}=7\tev$ recorded in 2010 with the ATLAS detector}\/},
  \href{http://dx.doi.org/10.1140/epjc/s10052-013-2306-0}{Eur.\ Phys.\ J.\ C
  {\bf 73}, 2306 (2013)},
\href{http://arxiv.org/abs/1210.6210}{{\tt arXiv:1210.6210 [hep-ex]}}.
%%CITATION = ARXIV:1210.6210;%%.

\end{thebibliography}\endgroup

%%%%%%%%%%%%%%%
% Authorlist
%%%%%%%%%%%%%%%
\onecolumn
\clearpage
% ATLAS Collaboration author list
% Data extracted on 26-Jan-2015 for paper reference TOPQ-2013-03
\begin{flushleft}
{\Large The ATLAS Collaboration}

\bigskip

G.~Aad$^{\rm 84}$,
B.~Abbott$^{\rm 112}$,
J.~Abdallah$^{\rm 152}$,
S.~Abdel~Khalek$^{\rm 116}$,
O.~Abdinov$^{\rm 11}$,
R.~Aben$^{\rm 106}$,
B.~Abi$^{\rm 113}$,
M.~Abolins$^{\rm 89}$,
O.S.~AbouZeid$^{\rm 159}$,
H.~Abramowicz$^{\rm 154}$
% ...
\bigskip
\\
% ...
$^{11}$ Institute of Physics, Azerbaijan Academy of Sciences, Baku, Azerbaijan\\
% ...
$^{84}$ CPPM, Aix-Marseille Universit{\'e} and CNRS/IN2P3, Marseille, France\\
% ...
$^{89}$ Department of Physics and Astronomy, Michigan State University, East Lansing MI, United States of America\\
% ...
$^{106}$ Nikhef National Institute for Subatomic Physics and University of Amsterdam, Amsterdam, Netherlands\\
% ...
$^{112}$ Homer L. Dodge Department of Physics and Astronomy, University of Oklahoma, Norman OK, United States of America\\
$^{113}$ Department of Physics, Oklahoma State University, Stillwater OK, United States of America\\
% ...
$^{116}$ LAL, Universit{\'e} Paris-Sud and CNRS/IN2P3, Orsay, France\\
% ...
$^{152}$ Institute of Physics, Academia Sinica, Taipei, Taiwan\\
% ...
$^{154}$ Raymond and Beverly Sackler School of Physics and Astronomy, Tel Aviv University, Tel Aviv, Israel\\
% ...
$^{159}$ Department of Physics, University of Toronto, Toronto ON, Canada\\
% ...
\end{flushleft}

\end{document}